\documentclass[preprint]{aastex}
\usepackage{graphicx}
\usepackage{amsmath}
\usepackage{gensymb}

\shorttitle{Finding Faint Asteroids}
\shortauthors{Heinze et al.}

\begin{document}

\title{Digital Tracking Observations Can Discover Asteroids Ten Times Fainter than Conventional Searches}

\author{Aren N. Heinze\altaffilmark{1,2}, Stanimir Metchev\altaffilmark{3,4,1}, and Joseph Trollo\altaffilmark{3}}

\altaffiltext{1}{Physics \& Astronomy Department, Stony Brook University, Stony Brook, NY 11794-3800, USA; aren.heinze@stonybrook.edu}
\altaffiltext{2}{Visiting astronomer, Kitt Peak National Observatory, National Optical Astronomy Observatory, which is operated by the Association of Universities for Research in Astronomy (AURA) under a cooperative agreement with the National Science Foundation.}
\altaffiltext{3}{Physics \& Astronomy Department, The University of Western Ontario, London, ON N6A 3K7, Canada; smetchev@uwo.ca}
\altaffiltext{4}{Centre for Planetary and Space Exploration, The University of Western Ontario, London, ON N6A 3K7, Canada; smetchev@uwo.ca}

\begin{abstract}
We describe digital tracking, a method for asteroid searches that greatly increases the sensitivity of a telescope to faint unknown asteroids. It has been previously used to detect faint Kuiper Belt objects using the Hubble Space Telescope and large ground-based instruments, and to find a small, fast-moving asteroid during a close approach to Earth.  We complement this earlier work by developing digital tracking methodology for detecting asteroids using large-format CCD imagers. We demonstrate that the technique enables the ground-based detection of large numbers of new faint asteroids.  Our methodology resolves or circumvents all major obstacles to the large-scale application of digital tracking for finding main belt and near-Earth asteroids. We find that for both asteroid populations, digital tracking can deliver a factor of ten improvement over conventional searches. Digital tracking has long been standard practice for deep Kuiper Belt surveys, but even there our methodology enables deeper integrations than have yet been attempted. Our search for main belt asteroids using a one-degree imager on the 0.9m WIYN telescope on Kitt Peak validates our methodology, delivers sensitivity to asteroids in a regime previously probed only with 4-meter and larger instruments, and leads to the detection of 156 previously unknown asteroids and 59 known objects in a single field. Digital tracking has the potential to revolutionize searches for faint moving objects ranging from the Kuiper Belt through main belt and near-Earth asteroids, and perhaps even anthropogenic space debris in low Earth orbit.
\end{abstract}


\section{Introduction}
Thousands of new asteroids are discovered every year by dedicated searches such as LINEAR \citep{Stokes2000}, Spacewatch \citep{Gehrels1996}, the Catalina Sky Survey \citep{Larson2007}, NEAT \citep{Helin1997,Pravdo1999}, and others; by more general sky surveys such as Pan-STARRS \citep{Denneau2013} and WISE \citep{mainzer2012}; and by advanced amateur astronomers.  The fundamental methodology of the detection is the same in all cases: an asteroid is identified as an object that changes its position in a systematic fashion relative to the starfield, and is detected individually in each of three to five different images \citep{Larson2007}.  A good example of a highly successful automated implementation of this methodology is the tracklet creation module of the Pan-STARRS Moving Object Processing System (MOPS; Denneau et al. 2013).

\begin{figure}
\includegraphics[scale=0.29]{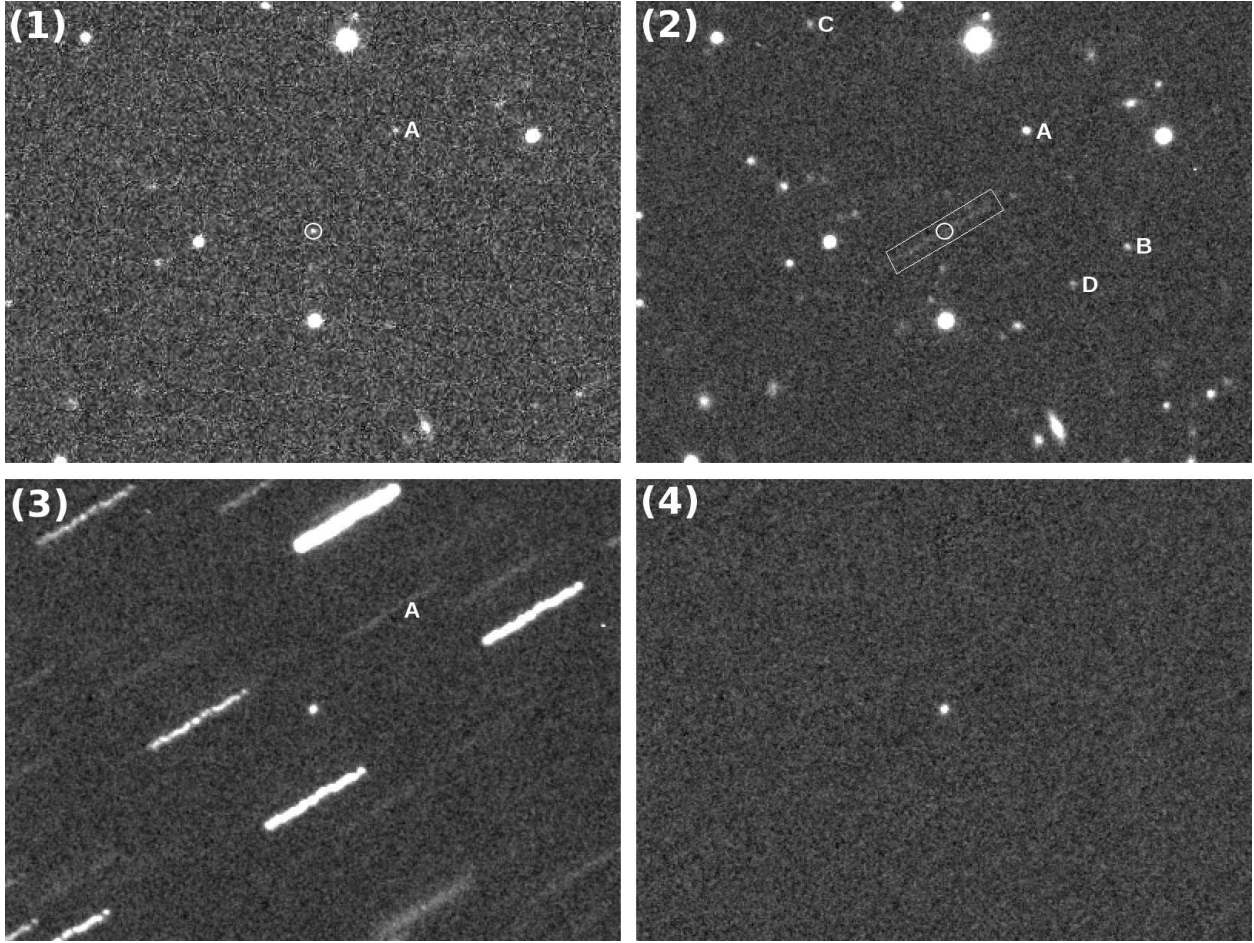}
\caption{Usefulness of digital tracking. \textbf{Panel 1:}
Single two-minute exposure showing a previously unknown asteroid (circled) and
similar-brightness Star A.  \textbf{Panel 2:} Median stack of
20 two-minute exposures registered on stars.  The asteroid has
blurred into a faint streak, while Star A and several fainter stars are
clearly seen. Rectangular vs. circular apertures around the asteroid 
illustrate two different (both unsatisfactory) ways of trying to detect the streak.
\textbf{Panel 3:} Stack of same 20 images
digitally tracked to register on the asteroid.  The asteroid is very bright, while
stars fainter than Star A leave undetectable streaks.
\textbf{Panel 4:} Like Panel 3 but with stars subtracted to
allow rapid automatic detection of asteroids.
\label{fig:trackexamp}}
\end{figure}

The fundamental methodology described above implies that an asteroid can only be discovered if it is detected above some noise threshold on each one of a series of individual images. For a given telescope and detector, sensitivity to faint stars increases with exposure time $t$, being proportional to $\sqrt{t}$ in the typical, background-limited case.  By contrast, sensitivity to faint asteroids generally ceases to increase for exposures longer than a maximum useful exposure time $\tau_M$, which is equal to the time an asteroid takes to move an angular distance corresponding to the resolution of the system being used to detect it.  On images with exposures longer than $\tau_M$, asteroids blur out into streaks that fade into the background noise (see Figure \ref{fig:trackexamp}).  For typical asteroid-search telescopes delivering 1--1.5 arcsecond resolution, $\tau_M$ is about 2 minutes for main-belt asteroids (MBAs) at opposition. Kuiper Belt objects (KBOs) move more slowly and allow longer exposures of up to 15 minutes; near-Earth objects (NEOs) can move so fast that $\tau_M$ gets as short as one second \citep{Shao2014}.  By contrast, 4--6 hour combined integration times are routinely used for imaging faint stars and galaxies using ground-based telescopes.  The maximum useful exposure time thus imposes a severe handicap on our sensitivity to NEOs, MBAs, and even KBOs. 

Digital tracking overcomes the handicap imposed by the maximum useful exposure time. It enables the detection of previously unknown moving objects using integrations lasting up to eight hours (i.e., one entire night) or even spanning several nights in the case of KBOs. With digital tracking, sensitivity to faint moving objects increases as the square root of the integration time just as it does for stars and galaxies. Digital tracking represents a fundamentally different detection paradigm from that employed in MOPS \citep{Denneau2013} and all other major asteroid surveys. We first became aware of the method's potential from reading the work of \citet{Bernstein2004}, who used it to detect extremely faint KBOs using long integrations with the Hubble Space Telescope.  More recently, \citet{Shao2014} and \citet{Zhai2014} have developed the technique in a different context: detection of NEOs during close approaches using a specialized CCD imager with very fast readtime. 

Such results have demonstrated the power of digital tracking, but they have only begun to
exploit its potential for discovering new Solar System objects.  We demonstrate this
by developing and applying digital tracking methods in a new context: the use of large format,
mosaic CCD imagers to detect hundreds of previously unknown asteroids in a single field. 
In Section \ref{sec:basic}
we describe the basics of digital tracking, and outline its capabilities and limitations.
We briefly review past applications of the method in Section \ref{sec:overview},
in Sections \ref{sec:Data}--\ref{sec:results} we report our new methodology and
results, and in Section \ref{sec:maincomp} we discuss the advantages and disadvantages
of digital tracking as compared to conventional surveys.  We conclude in Section \ref{sec:conc}
with a summary of our results and their implications for future surveys.



\section{How Digital Tracking Works} \label{sec:basic}

Here we will briefly outline the basics of digital tracking.  Although we will use the word `asteroid', the reader should take this as a generic term for a moving object in the sky, remembering that the technique is applicable not only to asteroids (both NEOs and MBAs) but also to KBOs, to comets, and to anthropogenic satellites or space debris. 

Digital tracking begins with the acquisition of a large number of images with individual exposure times $ \le \tau_M$.  These images should all target approximately the same position in the sky, and should be obtained consecutively (or, at least, all within a limited time span; see Section \ref{sec:skymotion}). We then parameterize the space of possible asteroid motions (which we will refer to as `angular motion phase space') in such a way that we can define a region that contains the motions of the asteroids of interest.  We search this region of angular motion phase space using a finely spaced grid of sampling points. Each point corresponds to a possible asteroid trajectory that is traced out in the time spanned by the acquisition of the images, and that is fully specified except for an arbitrary translation that allows it to begin or end anywhere in the field of view covered by the images.  For each of these sampling points in angular motion phase space, we create a separate, sigma-clipped median stack of shifted input images (a `trial stack'), where the shifts applied to each image are calculated to correctly register any asteroids moving on the trajectory corresponding to that sampling point. Such an asteroid will appear as a point of light on the stacked image, while stars, galaxies, and asteroids with different motions will blur into streaks and fade into the background\footnote{That is, in principle.  In practice, stars, galaxies, and other stationary celestial objects should be removed using image subtraction before creating the digital tracking stacks.}. The final step in digital tracking is therefore the identification of point sources on the trial stacks, each of which corresponds to an asteroid.  Accurate measurements of the position, motion, and brightness of such asteroids are natural products of the detection process.

\begin{figure} 
\includegraphics[scale=0.8]{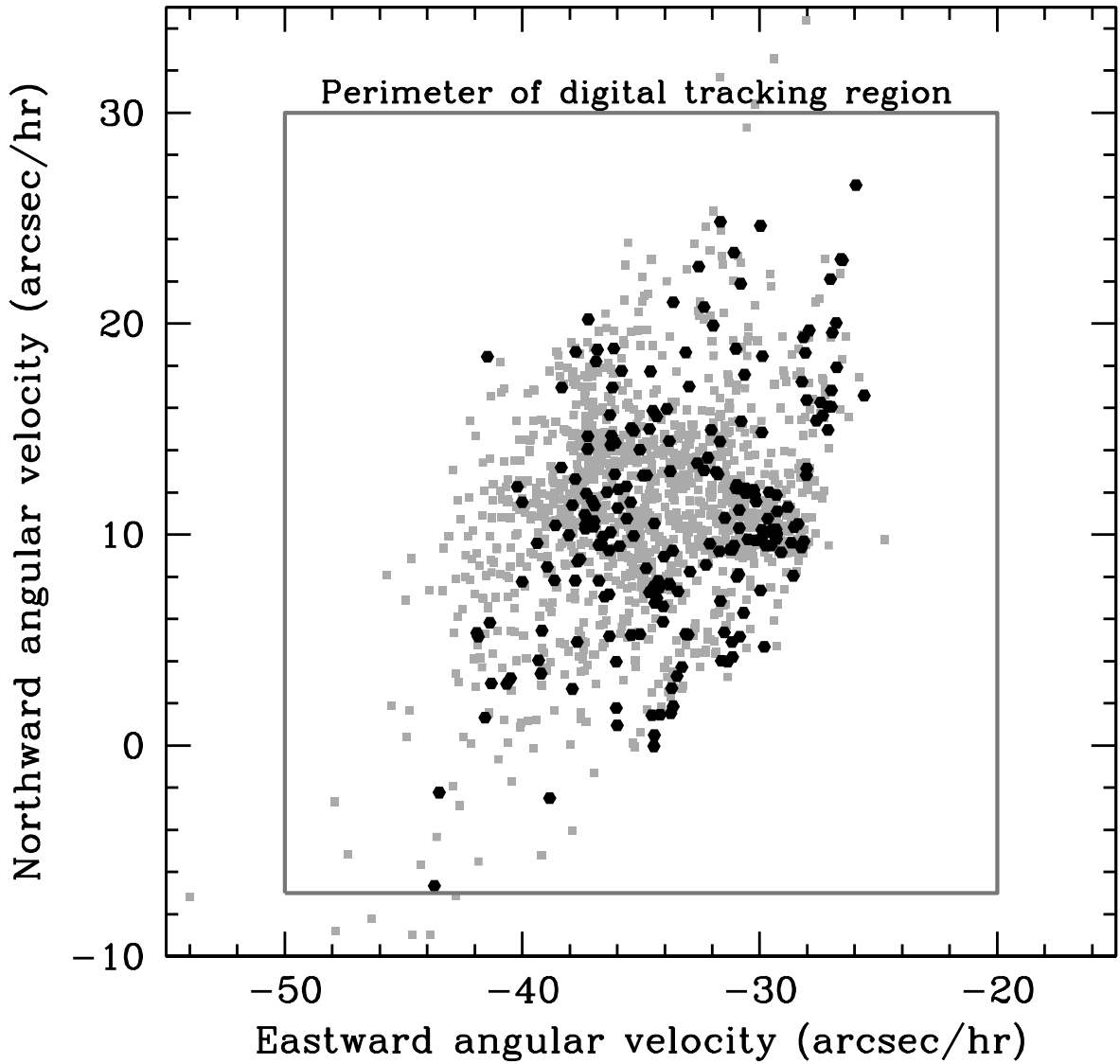}
\caption{Angular motion phase space for our observations on April 19, 2013. The measured angular velocities of asteroids detected in our data are shown as black hexagons.
The gray squares plotted to suggest the locus of main belt asteroids are known objects over a much
larger region of the sky (a disk 6 degrees in diameter) centered on our field at the time of our
observations. These motions were obtained from the Minor Planet Center. The rectangular outline
is the boundary of the region of angular motion phase space that we have searched.
Note that we have no detections in the corner regions far from the locus of known asteroids. This
provides evidence that our false positive rate is extremely low, as we further demonstrate
through the tests described in Sections \ref{sec:mancheck} and \ref{sec:falsepos}.
\label{fig:drift}}
\end{figure}

Two important facts render our specific implementation of digital tracking practical. First, the read noise of modern CCDs is negligible in the context of broadband optical imaging to detect asteroids, allowing us to stack large numbers of short-exposure images and obtain the same sensitivity as a single equivalent long exposure.  It is this property that makes digital tracking far superior to other ways of detecting faint asteroids (Section \ref{sec:noise}).  Second, for observations obtained from Earth's surface, the motions of asteroids and KBOs can be usefully approximated as linear motion in the plane of the sky, at constant angular velocity (Section \ref{sec:skymotion}). This allows us to use a simple two-dimensional parameterization of the angular motion phase space, as illustrated in Figure \ref{fig:drift}.

\subsection{Sensitivity of Various Detection Methods} \label{sec:noise}
We will now consider how the sensitivity of asteroid-search observations depends on integration time for different detection methods, and demonstrate that alternative methods to digital tracking do not overcome the sensitivity limit imposed by the maximum useful exposure. We note that \citet{Shao2014} have presented a similar analysis, illustrated especially by their Figures 5--6. 

We will refer to the cumulative exposure time of a stack of images as the \textit{integration time}, while we will use \textit{exposure time} to apply strictly to a single image. Thanks to the low read noise of modern CCDs, there is no significant noise penalty for dividing a long integration up into many shorter exposures. A digital stack of sixty images each taken with a one-minute exposure has the same sensitivity as a single image taken with a one-hour exposure. In fact, stacks of short exposures usually have superior sensitivity because they are not plagued by artifacts from severely saturated stars, and because cosmic rays can be rejected by sigma-clipping.  The dominant source of noise in broad-band images with astronomical CCDs is the Poisson noise of the sky background.  Let the sky background have a surface brightness of $F_{sky}$ in photons/s/arcsec$^2$, while a point source (star or asteroid) has a brightness of $F_{source}$ in photons/s.  The source is to be detected based on its flux within an aperture of radius $r$ in arcseconds.  The optimal value of $r$ for the detection and accurate measurement of faint sources is usually comparable to half the full width at half-maximum (FWHM) of the telescope's point spread function (PSF). Note that $r$ remains a meaningful concept even when sources are being detected using using PSF profile-fitting rather than aperture photometry: the detection is still effectively performed within a roughly circular region corresponding to the bright core of the PSF.  In either case, where readnoise and dark current are negligible, the signal-to-noise (SNR) level at which effectively stationary objects are detected in integration lasting $t$ seconds is given by:

\begin{equation}
SNR(t) = \frac{t F_{source}}{\sqrt{\pi r^2 t F_{sky}}} \propto \sqrt{t}
\label{eq:snr01}
\end{equation}

This equation describes the detection of stars on Panel 1 of Figure \ref{fig:trackexamp}, and of the asteroid since two minutes does not exceed the maximum useful exposure time $\tau_M$. Note that the circles drawn around the asteroid in the figure do not correspond to the optimal photometric radius $r$: they are exaggerated for clarity.  Equation  \ref{eq:snr01} accurately predicts that on Panel 2, where 20 images are stacked for an effective 40-minute integration, the SNR of stellar images (or, equivalently, the minimum stellar flux that can be detected) is improved by a factor of $\sqrt{20}$.  The same is not true of the moving asteroid, which is elongated into a streak on this panel.

The circular aperture in Panel 2 illustrates how one could measure the SNR of the asteroid's streaked image within an aperture with the same optimal radius $r$ used to detect un-blurred point sources.  The SNR for detecting the asteroid in this way is:

\begin{equation}
SNR(t) = \frac{\tau_M F_{source}}{\sqrt{\pi r^2 t F_{sky}}} \propto t^{-1/2}
\label{eq:snr02}
\end{equation}

\noindent for $t > \tau_M$. Equation \ref{eq:snr02} accurately predicts that for $t > \tau_M$, the SNR of any given PSF-sized portion of the asteroid trail actually \textit{decreases} with increasing integration time.  This is seen in Figure \ref{fig:trackexamp}: the streak in Panel 2 is a less obvious signal than the point source in Panel 1.  However, Panel 2 also illustrates an alternative way of measuring the asteroid: one could use an elongated aperture with width 2$r$ and length $(t/\tau_M) 2r$ to encompass the whole streak.  The SNR of this type of measurement is:

\begin{equation}
SNR(t) = \frac{t F_{source}}{\sqrt{4 r^2 (t/\tau_M) t F_{sky}}} \propto t^{0}.
\label{eq:snr03}
\end{equation}

Since the SNR does not decrease with increasing integration time, this method is an improvement on that described by Equation \ref{eq:snr02}. It is the best way of detecting asteroids on images with exposures longer than $\tau_M$. However, it still indicates that asteroids, unlike stars, cannot be detected with increasing sensitivity using exposures of increasing length. By contrast, digital tracking restores the $\sqrt{t}$ dependence of Equation \ref{eq:snr01} for moving objects on integrations with lengths of up to hundreds of times $\tau_M$.

\subsection{The Approximation of Linear Motion at Constant Velocity} \label{sec:skymotion}

Digital tracking is possible even for objects that exhibit nonlinear sky motions \citep{Bernstein2004}, but for such objects it becomes more computationally challenging, especially for the high data volumes produced by large format CCD imagers.  To keep the analysis computationally tractable, our current implementation uses the approximation of linear motion at a constant velocity. This means that the angular motion phase space we search has only two dimensions, which we parameterize in terms of on-sky angular velocities in right ascension (RA) and declination (DEC), as illustrated by Figure \ref{fig:drift}. The price of simplifying the angular motion phase space in this way comes in the form of a maximum duration $\tau_{lin}$ for digital tracking integrations targeting a given population of objects. The limit is reached when the on-sky tracks of the target objects deviate from the best-fit linear, constant velocity approximation by one resolution element (e.g., one arcsecond). Thus, while the maximum useful individual exposure $\tau_M$ is set by the angular velocity itself, $\tau_{lin}$ is set by the first time derivative of the angular velocity, corresponding to curvature and/or acceleration in the sky motion of the target object. As we determine below, $\tau_{lin}$ is large relative to $\tau_M$ for all target populations. The linear, constant velocity approximation is sufficient for very long, sensitive digital tracking integrations.

\begin{figure} 
\plottwo{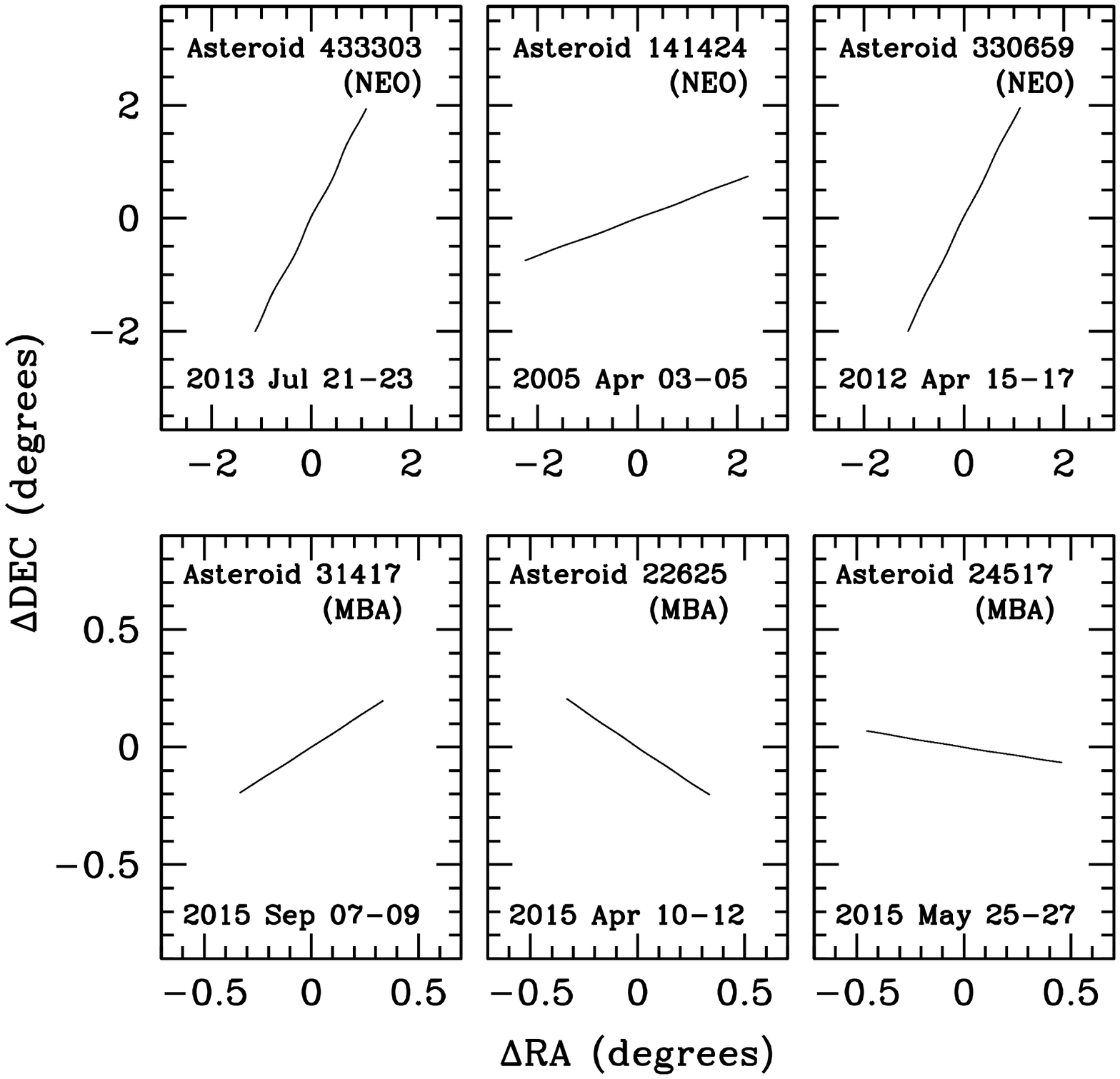}{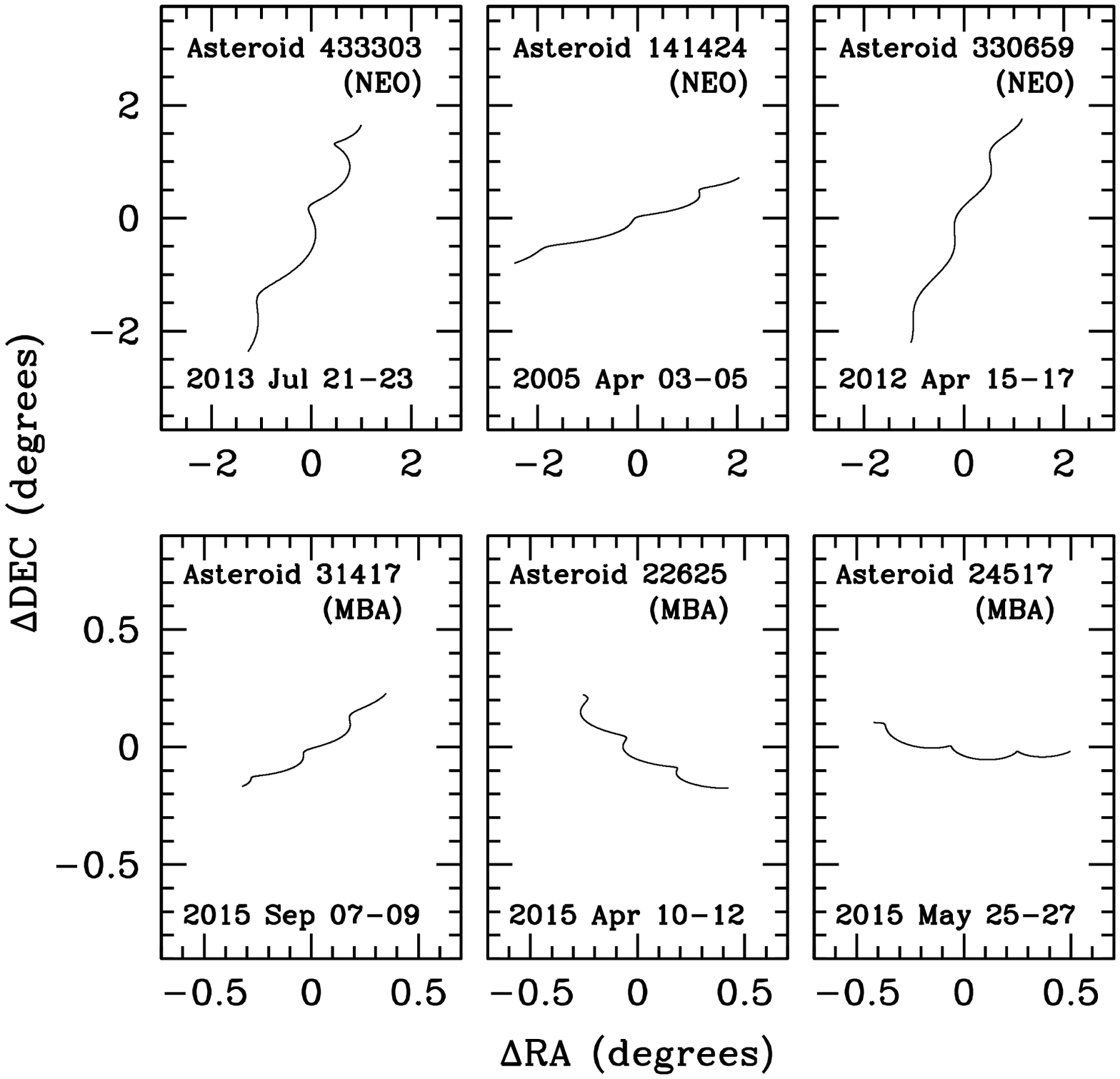}
\caption{The Earth's rotation causes the angular velocities of asteroids observed
from non-arctic latitudes to oscillate with a period of one day, causing a waviness
in their on-sky tracks.
\textbf{Left:} On-sky tracks of six representative
asteroids over three days, as calculated by JPL's Horizons system for
an observer located at Cerro Tololo, Chile. The tracks are plotted even through
intervals when the asteroids would be
below the horizon, as if they could be continuously observed through
a transparent Earth.
\textbf{Right:} Same as left, but with the nonlinear component of the
motions amplified (10$\times$ for NEOs and 25$\times$ for MBAs) to show
the effect of Earth's rotation. Distances here are 0.1--0.15 AU for the NEOs and
1.1--1.5 AU for the MBAs. Since the amplitude of the
nonlinear effects scales with inverse distance, it is about 10 times larger for the NEOs --- hence
the different plot scales and amplification factors.
\label{fig:trochoid}}
\end{figure}

\begin{figure} 
\plottwo{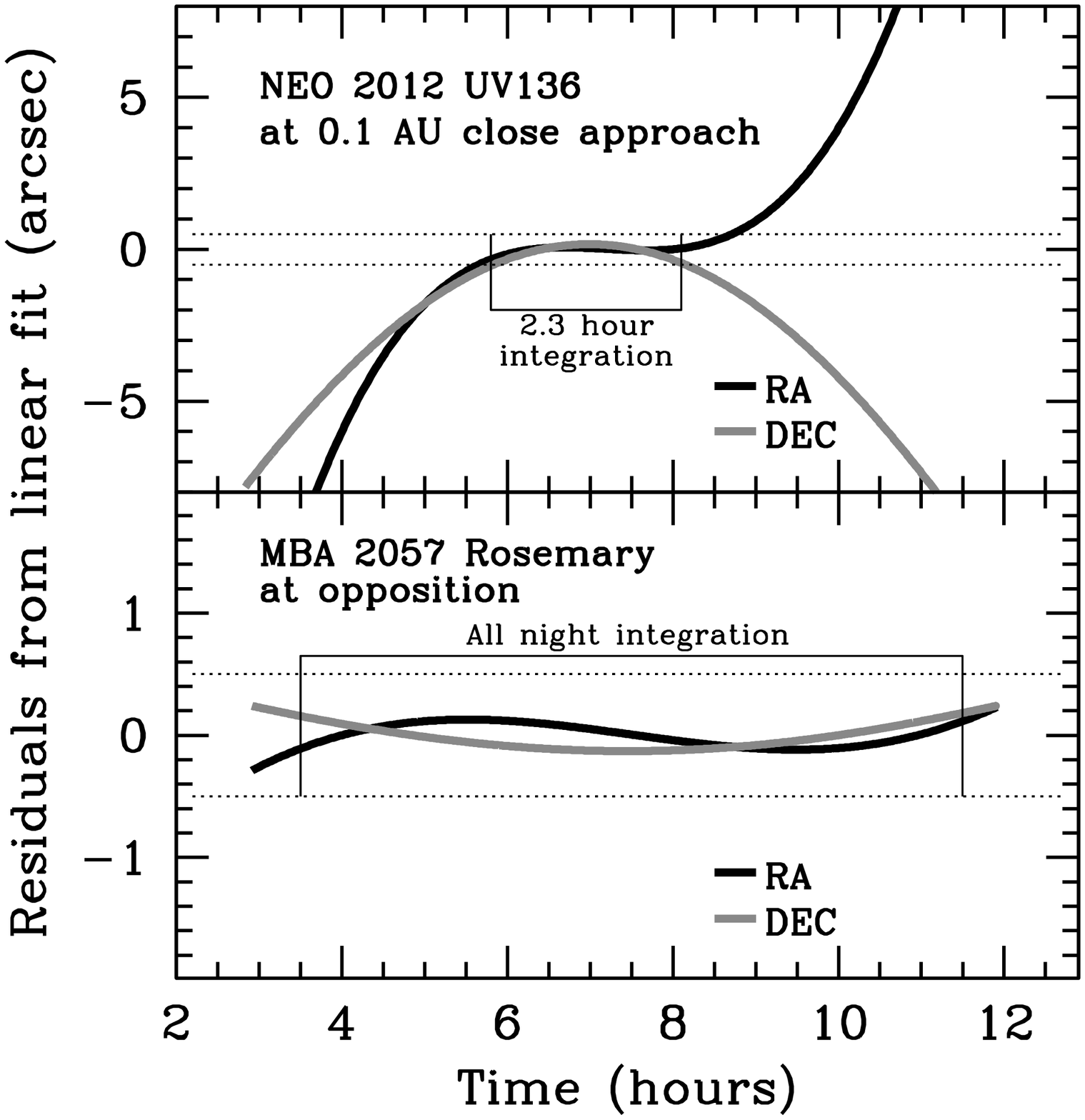}{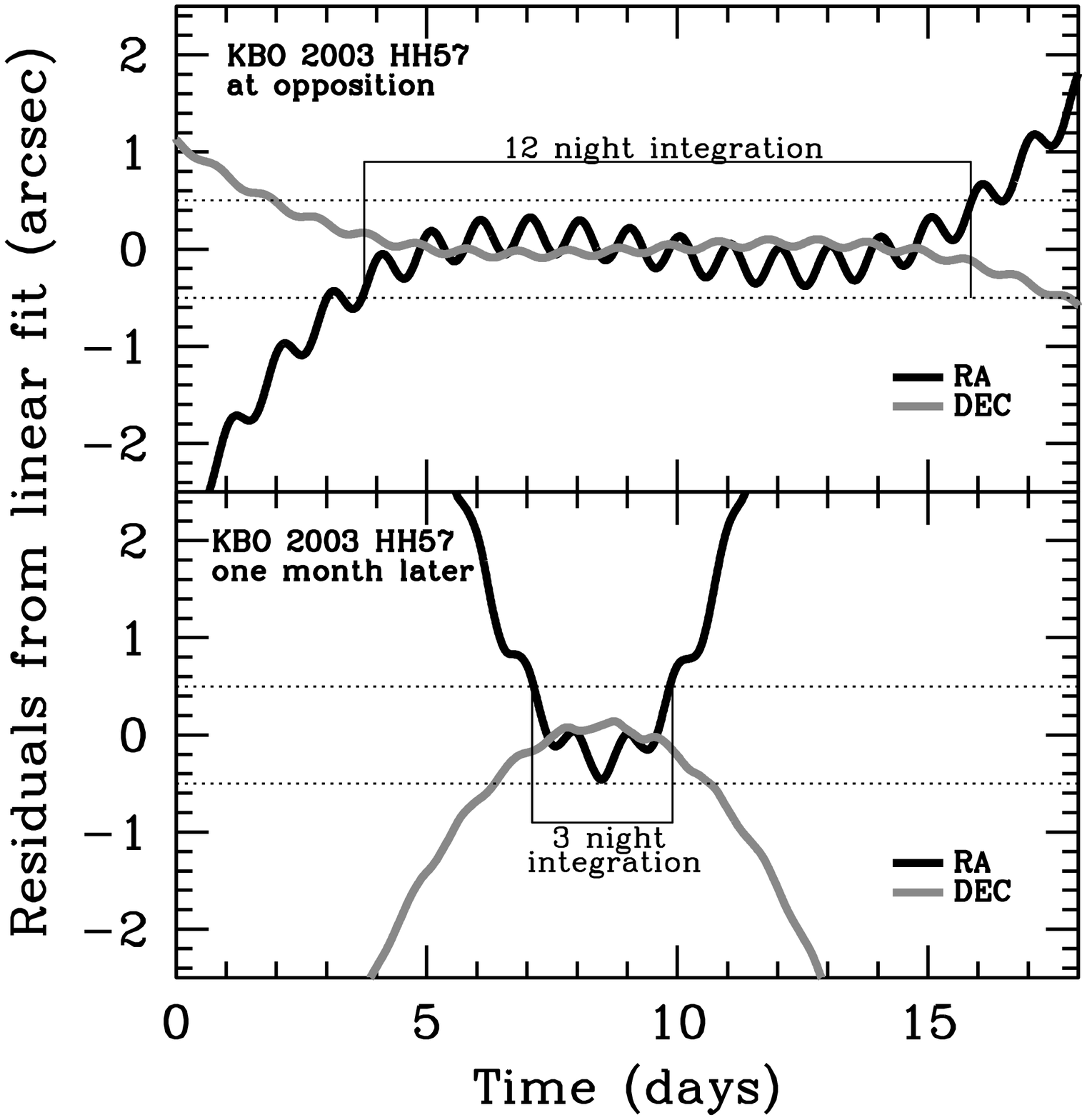}
\caption{Deviations of sky motions from the linear trajectory at constant
angular velocity that we use to approximate objects' sky motions.  The objects
shown are representative examples of their respective classes.  Topocentric positions were
obtained for Kitt Peak from the JPL Horizons system. Dotted lines give the $\pm 0.5$ arcsec
range of acceptable deviation during a digital-tracking integration. The time
an object stays between these lines defines $\tau_{lin}$, the maximum integration time
discussed in Section \ref{sec:skymotion}.
\textbf{Top Left:} For a near-Earth asteroid at a geocentric distance of
0.1 AU, a two-hour integration is possible if centered on zero hour angle, as here.
\textbf{Bottom Left:} For main belt asteroids, 8--9 hour integrations are always
possible. \textbf{Right:} For Kuiper Belt objects near opposition, integrations
of up to twelve nights are possible. One month away from opposition, $\tau_{lin}$ drops to three nights.
\label{fig:skymotions}}
\end{figure}

For NEOs and MBAs, the dominant cause of acceleration and curvature in on-sky tracks is the Earth's rotation. This imposes a 24-hr sinusoidal variation in the asteroid's on-sky velocity, and causes a waviness its trajectory with the same period (Figure \ref{fig:trochoid}). The amplitudes of the acceleration and curvature due to Earth's rotation are inversely dependent on the asteroid's distance from the Earth, since they are angular effects produced by the observer's motion. For MBAs, the acceleration and curvature are weak enough that $\tau_{lin}$ is always longer than the $\sim 8$ hr limit imposed on digital tracking integrations by the ordinary observing considerations of twilight and high airmass\footnote{Note, however, that $\tau_{lin}$ for MBAs is always less than 24 hours: digital tracking integrations spanning multiple nights are not possible for these objects.}. Eight-hour integrations are therefore possible on MBAs with residuals considerably less than one arcsecond (Figure \ref{fig:skymotions}), and the maximum integration time for these objects will be set by twilight or airmass rather than $\tau_{lin}$. This is true for MBAs regardless of viewing geometry: it applies equally to objects at opposition and far from opposition.

All-night integrations are not possible for NEOs making close approaches to the Earth, but 1--2 hr integrations work well up to very close distances. The acceleration and curvature induced by Earth's rotation both reach a stationary point when the asteroid transits the observer's meridian: that is, at zero hour angle.  Thus, digital tracking integrations on NEOs can be longer if the observations are centered on zero hour angle.  For a closely-approaching NEO at a distance of 0.1 AU from Earth, a two hour integration is possible if centered on zero hour angle, as shown in Figure \ref{fig:skymotions}. For observations far from the meridian, $\tau_{lin}$ drops to about one hour.

KBOs are sufficiently distant that the diurnal oscillation imposed on their tracks by Earth's rotation has an amplitude considerably less than 1 arcsecond, as shown in Figure \ref{fig:skymotions}. The dominant cause of acceleration and curvature in the on-sky tracks of KBOs is therefore the curvature of Earth's orbit. It follows that $\tau_{lin}$ is greater than 24 hours: digital tracking integrations spanning multiple nights are possible. The effects from Earth's orbit reach a stationary point for objects exactly at opposition, so KBO integrations centered on opposition can last up to twelve nights (Figure \ref{fig:skymotions}).  One month after opposition, $\tau_{lin}$ shortens to three nights. 

To sum up, the $\tau_{lin}$ values for NEOs, MBAs, and KBOs are all hundreds of times longer than $\tau_M$ for the corresponding objects.  Since the sensitivity gain from digital tracking observations is a factor of $\sqrt{t/\tau_M}$, digital-tracking observations obeying the requirement that $t < \tau_{lin}$ can detect objects at least ten times fainter than conventional searches with the same telescope and imager at the same SNR level.

\subsection{Numbers of Trial Vectors to Search for MBAs, NEOs, and KBOs} \label{sec:points}
The computational challenge of digital tracking depends on the number of trial
points in angular motion phase space that must be probed in the analysis.  In the context
of our approximation of linear, constant-velocity motion, each point in
angular motion phase space corresponds to a two dimensional vector; hence we will also
refer to them as trial vectors. We have already introduced the term `trial stack'
for the image stack corresponding to a given trial vector. 
The required number of trial vectors is
determined by the size of the region in angular motion phase space that is to be 
explored and the maximum
permissible spacing between sampling points (vectors) in this region. 
We consider the spacing of the sampling grid and the size of the region of
angular motion phase space in turn. We restrict the current discussion to the case of
linear motion at a constant velocity as described in Section \ref{sec:skymotion}.

Trial vectors in the resulting two-dimensional angular motion phase space
should be spaced finely enough that every object will be imaged in at least
one trial stack with a blur length less than some maximum acceptable
value $b_{max}$, which will normally be about one arcsecond.  Let the
grid spacing in angular motion phase space be $\Delta_m$ and
the total time spanned by the integration be
$t_{int}$ (note that observational overheads always make this somewhat longer
than the cumulative integration time $t$). The worst possible blur will be for an object centered between
four grid points. Its blur length on each of the four corresponding trial stacks
is identical and is equal to $t_{int} \Delta_m/\sqrt{2}$. The optimal grid spacing
is therefore:

\begin{equation}
\Delta_m  = \frac{\sqrt{2} \; b_{max}}{t_{int}}
\label{eq:blur}
\end{equation}

While the grid spacing depends only on the temporal span $t_{int}$ of the digital
tracking integration, the size of the grid that should be explored depends
on the type of object being sought.  We will now consider the requirements
for targeting MBAs, NEOs, and KBOs.

\subsubsection{Trial Vectors for Main Belt Asteroids} \label{sec:mbatrial}
MBAs near opposition are in their retrograde loops
and thus have negative (westward) motions in RA, with speeds between 20 and 50 arcsec/hr.
Their motions in DEC depend on their orbital inclinations and the trendline of the ecliptic
(e.g., the DEC motions skew northward near 12:00 RA, southward at 00:00, and have mean
zero at 06:00 or 18:00).  In any case, a range of $\pm16$ arcsec/hr centered on the mean value
encompasses most of them (however, as illustrated by Figure \ref{fig:drift}, we choose a slightly larger north-south range when analyzing our April 2013 observations, to increase the chance of including interesting outliers). 
For a more typical north-south range, the rectangular region of angular motion phase space to 
be explored has a size of about 30$\times$32 arcsec/hr.
If we consider an eight-hour digital tracking integration, with $b_{max}=1$ arcsec,
the grid spacing $\Delta_m$ is 0.18 arcsec/hr and the number of points required to
span the grid is 31,000.  Where each point requires the stacking of a few hundred
large-format CCD images, the computational demand is considerable --- but nonetheless
feasible even with modern desktop workstations.

As an example of the computational requirements, the April 2013 test survey we describe in
Sections  \ref{sec:Data}--\ref{sec:results} consists of two data sets of 126 and 130 images.
After processing, each image is 12,000 pixels square and thus totals 144 megapixels, where each
pixel is 32 bits.
Our digital tracking analysis probed 28,086 distinct trial vectors for each data set (equivalently,
112,344 trial vectors per data set with the images split into quadrants as described in Section \ref{sec:digipars}). 
We can quantify the processing task in terms of `vector-pixels',
where the number of vector-pixels is defined as the number of input images times the number of pixels
per image times the number of trial vectors. Our test survey probed $1.035\times10^{15}$ vector-pixels (i.e. $1.44\times10^8$ pixels per image times 256 total images times 28,086 trial vectors). Using a 32-core desktop workstation,
our processing rate was $8.1\times 10^{11}$
vector-pixels per hour, allowing us to process the entire two-night survey in about 50 days. 
For comparison, the Tightly Coupled System (TCS) supercomputer at the SciNet HPC
Consortium\footnote{http://www.scinethpc.ca/} (funded by the Canada Foundation for Innovation), to which we have access for future work, has well over 100 times the processing power of our desktop
and could re-analyze the entire two-night survey in only a few hours.

\subsubsection{Trial Vectors for Near-Earth Objects: A Statistical Survey} \label{sec:neostat} 
We will present two examples of hypothetical digital tracking surveys targeting NEOs
within 0.25 AU of the Earth. Both surveys involve digital tracking integrations lasting
one hour or less, as is appropriate for NEOs (Section \ref{sec:skymotion}), and hence more
than one digital tracking integration can be obtained each night. 

The first hypothetical NEO survey is presented in this Section, and is aimed at detecting
and calculating approximate orbits for extremely small NEOs to probe their population statistics.
Most of these objects will be too faint for extensive followup observations, which is acceptable
because the 2--4 day span of the survey itself will yield orbits sufficiently accurate for
statistical analysis. Also, the faint objects will be too small to pose a significant impact risk to humans
on Earth, so long-term followup observations to predict impacts or retire impact risk are unnecessary. 
A second hypothetical survey, this time focused on risk retirement, is described in Section \ref{sec:risk}.

NEOs can have a wide range of possible motions depending on their distances
and Earth-approach geometries. Our hypothetical statistical survey will target asteroids
close to Earth, at distances between 0.1 and 0.25 AU. This is a challenging case due to
the fast sky velocities of such closely-approaching objects, and therefore it
provides a good example both of the limitations and the power of digital tracking. 

We have obtained ephemerides for known NEOs from the Minor Planet Center (MPC) to determine the
appropriate range of angular motion phase space for our hypothetical survey. Using the
full list of 1,915 numbered NEOs (Amors, Atens, and Apollos) given by the MPC as of June 2015,
we search for Earth encounters at distances between 0.1 and 0.25 AU where the celestial coordinates
of the asteroids involved are within 20$^{\circ}$ of the antisolar point\footnote{As the antisolar point is 180$^{\circ}$ from the
sun by definition, we apply this constraint to the MPC ephemerides by requiring
solar elongation $\ge 160^{\circ}$.}. We confine
our search to the years 2000 through 2020, and use an ephemeris sampling interval of
one week. We find 699 encounters, involving 246 distinct asteroids, which satisfy our criteria.
In this sample of 699 encounters, the median on-sky velocity of the asteroids is 140 arcsec/hr and
the 90th percentile velocity is 340 arcsec/hr. We note that if we do not confine the sample
to encounters where the asteroid is near the antisolar point, the median and 90th percentile angular velocities go up to 
200 arcsec/hr and 460 arcsec/hr, respectively. We suspect this is due to the inclusion of more
high-inclination objects that are rarely found near the ecliptic or the antisolar point, and hence
are less likely to appear in the initial sample. A target field near
the antisolar point (hence detecting asteroids that are near opposition) is strongly to be preferred
for digital tracking observations, because asteroid phase functions and observability constraints
result in considerably lower sensitivity with any other targeting. The angular velocities from the
sample of encounters that is restricted to large solar elongation are therefore more representative of those
in a real digital-tracking survey of NEOs, and we will use them to calculate the number of
trial vectors.

The maximum useful exposure
($\tau_M$) is typically about 30 seconds for objects moving at 140 arcsec/hr (given 1.2 arcsecond seeing),
and 13 seconds at 340 arcsec/hr. The read times for the current generation of large-format
imagers (e.g. 20 seconds for DECam at Cerro Tololo; Flaugher et al. 2012) render observations with exposures
much shorter than 30 seconds quite inefficient due to readout overhead. Thus, rather
than adopting an exposure time that is less than or equal to $\tau_M$ for all objects
in our target sample, our hypothetical NEO survey must
use an exposure time that is a compromise between inefficiency due to readout overhead
on the one hand, and reduced sensitivity due to trailing losses on the other.  Even
for fast-moving objects whose trailing losses are considerable, digital tracking
will still yield a large increase in sensitivity over conventional methods. Given
an exposure time of 30 seconds (less than $\tau_M$ for half the target population),
trailing losses will amount to only about 0.6 magnitudes for objects moving
at 340 arcsec/hr. This is calculated by comparing Equations \ref{eq:snr01} and \ref{eq:snr03}: 
the loss in SNR due to trailing is a factor of $\sqrt{(4/\pi)(t/\tau_M)}$. In the present case
we have $\sqrt{(4/\pi)(30/13)}$ = 1.71, which is 0.59 magnitudes.

The digital tracking analysis for our hypothetical, statistical NEO survey should then
search a region of motion space described by a circular 
disk of radius 340 arcsec/hr centered on the origin. Such a search
will motion-match 90\% of NEOs at geocentric distances between 0.1 AU and 0.25 AU,
and will also deliver sensitivity to slower-moving objects at even smaller distances. Given a one-hour
digital tracking integration with $b_{max}=1$ arcsec and thus $\Delta_m$ equal to 1.4 arcsec/hr, 185,000
trial points are required. Given 120 images of 100 megapixels each, the computational
requirement is $2.22 \times 10^{15}$ vector-pixels, and would take four months with
our multi-core desktop but only about 24 hours with SciNet's TCS supercomputer. This is for a
digital tracking integration lasting only one hour. Since several such integrations could
be obtained each night, and orbital statistics might require two or more nights, access to
a supercomputer is a practical necessity for a statistical survey of this type.

Future generations of large-format CCD imagers will have shorter readout times and may thus enable
digital tracking searches to use very short (1--5 sec) exposures and probe faster-moving NEOs for which 
$\tau_M$ is only a few seconds \citep{Shao2014}. However, targeting faster moving objects with shorter
exposure times greatly increases computational requirements, if the length of the digital tracking
integration is held fixed. Using a small-format but fast-readout CCD, Zhai et al. (2014) have
demonstrated that interesting sensitivity regimes
may be probed for very fast-moving objects even with quite short digital tracking integrations.
The potential sensitivity of long (e.g. one-hour) digital tracking integrations such as we have proposed
in this section will remain even greater, and we may hope that supercomputer capabilities will
continue to increase in parallel with the development of next-generation CCD imagers with large
formats and fast readtimes. If computer capabilities increase sufficiently, long digital tracking integrations targeting
very fast-moving NEOs may be computationally feasible by the time large-format
detectors capable of observing with the required very short exposures are available.

Digital tracking analyses requiring days or even weeks of computer time to process each night's data are acceptable
for surveys aimed at probing the statistics of an asteroid population. The science goals of
such a survey do not require followup imaging by other observatories, nor the calculation of
sufficiently precise orbits to allow long-term recovery observations or impact prediction.
Such surveys will, however, generally make at least a serendipitous contribution to long-term
orbit calculation for some objects, by producing recoveries (or precoveries) at relatively
large distances of objects that were (or will be) discovered by other, less sensitive surveys
during closer approaches to Earth. To reap the benefits of such detections, it is very desirable
that all asteroid detections should be reported to the MPC, even in the case of a statistical survey.

\subsubsection{Trial Vectors for Near-Earth Objects: A Risk-Retirement Survey} \label{sec:risk} 
The processing times discussed in Section \ref{sec:neostat} above are too long for an NEO
survey aimed at calculating precise orbits, making long-term impact predictions, and retiring terrestrial
impact risk. Such surveys should run continuously and process all data promptly in order to give
other observers the opportunity to followup potentially hazardous asteroids and refine their orbits.
The data from each night should therefore be processed before the next night's observations.
Several surveys of this type are currently underway (for example, Gehrels \& Jedicke 1996; Helin et al. 1997; Pravdo et al. 1999; Stokes et al. 2000; Larson 2007),
but none use digital tracking at present. We will now explore what role digital tracking could play, if any,
in such a survey.

We will consider a system like that of the Catalina Sky Survey,
which uses a 16 megapixel CCD to cover an 8.2 square-degree field of view with 2.6-arcsecond pixels,
with a standard exposure time of 30 seconds \citep{Larson2007}. Although the acquisition of a dedicated
supercomputer would be a reasonable step for a survey of this type transitioning to a digital tracking
mode, we will assume that only a more modest processing capacity equal to twelve times that of our own desktop
workstation is available. Twelve such computers would cost about about \$60,000 in 2015, and would enable the processing of
$10^{13}$ vector-pixels per hour.

Like that described in Section \ref{sec:neostat}, our hypothetical risk-retirement survey 
will use digital tracking integrations spanning one hour or less, and will acquire several integrations
per night.  However, while the integration length of one hour used in Section \ref{sec:neostat} was
set by $\tau_{lin}$ (see Section \ref{sec:skymotion}), here the integration length will be set by the limits
on processing power. We will assume that 8 hours of useful data are acquired each night
and that computer processing of this data runs continuously 24 hours each day. Thus, 24 hours
are available to process each 8 hours of data, and processing of each digital tracking integration
must take no longer than three times as long as data acquisition. 

We will now attempt to determine an optimum value for the length of an individual integration. Specifically,
we will seek to optimize $n$, the number of images per digital tracking integration.
The larger the number of images, the longer the integration and the greater the sensitivity (which scales
as $\sqrt n$). On the other hand, a longer integration time $t_{int}$ also requires a finer sampling
interval $\Delta_m$ for trial vectors in angular motion phase space (Equation \ref{eq:blur}). A finer
sampling interval means the available compute power is only sufficient to cover a smaller region of angular
motion phase space, which means we will detect a smaller fraction of the target asteroids. Additionally,
a longer integration time $t_{int}$ on each field means fewer fields can be covered each night.

We find that if the readout time and other overheads are negligible, $n=40$ appears as a good compromise.
This increases the sensitivity by a factor of 6.3 (2.0 magnitudes) relative to single exposures, and
the digital tracking analysis could motion-match over half the asteroids
within 0.05 AU and 90\% of those between 0.05 and 0.1 AU. These statistics are calculated using ephemerides
from the MPC for 1,098 close encounters of known asteroids between 2000 and 2020, and unlike those used
in Section \ref{sec:neostat} they are not restricted to encounters where the asteroid is near the anti-solar point.

The 40 images, each with a 30-second exposure, would be taken over a time interval of 20 minutes. As each image has 16 megapixels,
the total number of pixels is $6.4 \times 10^8$. Three times the integration time, or 1 hour, is available to
process the data, meaning that $10^{13}$ vector-pixels may be processed and thus $10^{13}/6.4 \times 10^8 = 15,625$
trial vectors may be probed. Given the 2.6-arcsecond pixels of the detector, we set the allowable blur 
$b_{max}$ to 3.0 arcsec. Equation \ref{eq:blur} then gives a sampling interval $\Delta_m$ of 14.2 arcsec/hr.
Given this sampling interval, with 15,625 trial vectors we can probe a circular region of radius 1,000 arcsec/hr
in angular motion phase space. Centering this region on the origin, it follows that all asteroids with
angular velocities slower than 1,000 arcsec/hr can be detected. Based on the statistics of known NEOs,
half of all asteroids within 0.05 AU of Earth; 90\% of those between 0.05 and 0.1 AU, and 99\% of those
between 0.1 and 0.2 AU are moving slower than this.

To facilitate the calculation of orbits and confirm lower-significance detections, our hypothetical
NEO survey should obtain at least two digital tracking integrations of each field per night: that is,
80 images per field per night with the parameters described thus far.  The Catalina Sky Survey currently acquires only 
four images per field per night. Transitioning to a digital tracking mode would therefore reduce the
survey's nightly sky coverage by a factor of twenty. On the other hand, it would create sensitivity to a large
number of asteroids that are currently completely undetectable. As more and more of the
large NEOs are discovered and their impact risk is retired through accurate orbit calculation, digital tracking
may present an avenue for current surveys to go on detecting new objects which, though smaller than the
surveys' original targets, remain large enough to pose a regional impact hazard to humans on Earth.

The survey parameters described above could be adjusted to accommodate non-negligible readout overheads and/or a need to detect
faster-moving objects.  For example, if the readout overhead were 10 seconds, 34 images could be acquired over
a slightly longer interval of 22.7 minutes, and all asteroids with angular speeds below 1,000 arcsec/hr could
still be detected in the 68.1 minute processing time available for the digital tracking analysis. The sensitivity
increase relative to single exposures would be $\sqrt 34 = 5.8$, or 1.9 magnitudes. Alternatively,
to detect all asteroids with motions up to 2,000 arcsec/hr (a threshold which includes 88\% of those within 0.05 AU), we could stack
just 17 images, and the sensitivity increase would be a factor of 4.1 (1.5 magnitudes). 

Asteroids moving at 2,000 arcsec/hr will be 17-arcsecond streaks on individual 30-second exposures. With
2.6 arcsecond pixels, this corresponds to a sensitivity reduction of about 1.1 magnitudes due to trailing losses\footnote{We assume here that the effective resolution is approximately equal to the pixel scale of 2.6 arcsec, and thus $\tau_M$ for an object
moving at 2,000 arcsec/hr is 2.6/2,000 = 0.0013 hr  = 4.7 seconds.}, regardless
of whether digital tracking is used. New-generation imagers with very short readout times will allow efficient
observations with exposures much shorter than 30 seconds. These will be able to detect faster-moving asteroids
with reduced trailing losses, at the expense of increased computational requirements for digital tracking integrations. 
As we indicated in Section \ref{sec:neostat}, it is possible that the development of faster
computers will keep pace with that of fast-readout astronomical imagers, and thus
cutting-edge digital tracking integrations will become more rather than less tractable over time.
A dedicated supercomputer may still be required for a risk-retirement survey seeking
to push to the limits of detection for faint, fast-moving NEOs.

\subsubsection{Trial Vectors for Kuiper Belt Objects} 

The sky motions of KBOs are slow and are dominated not by their
own space velocities but by the much faster orbital motion of the Earth.  The region to be
searched is thus quite small. If we seek objects ranging from 30 to 100 AU
in geocentric distance, in orbits of arbitrary eccentricity with (non-retrograde)
inclinations up to $90 \degree$, the motions of all target objects in this
very generous dynamical range can still be enclosed in a rectangular region
of motion space with dimensions 3.7 arcsec/hr by 1.3 arcsec/hr (see the derivation
in Parker and Kavelaars 2010). For a three-night integration
with $b_{max}=1$ arcsec, only 8,000 points are required to search this
region.  Even the twelve-night integration we found to be possible in Section \ref{sec:skymotion}
would require just 180,000 points. While this is comparable to the number of trial vectors
required for one of the NEO integrations described in Section \ref{sec:neostat}, the NEO
integrations spanned one hour, as compared to 12 days for the Kuiper Belt. Thus, the computational requirements are relatively modest
even for the most ambitious digital tracking integrations targeting KBOs.

\subsection{Comparing Digital Tracking Parameters for NEOs, MBAs, and KBOs} \label{sec:comp3} 

We have considered digital tracking as applied to three very different target
populations: NEOs, MBAs, and KBOs, listed here in order of decreasing angular
velocity. In Section \ref{sec:skymotion} we
calculated $\tau_{lin}$, the maximum duration of a digital tracking integration using
the approximation of linear motion at constant angular velocity. We found that
for all three populations, $\tau_{lin}$ is hundreds of times longer than the maximum useful
exposure $\tau_M$ for single images. Thus, digital tracking offers a potential
sensitivity increase of at least a factor of 10 for all three populations. 
In Section \ref{sec:points}, we found that the limits on digital tracking integration
lengths imposed by $\tau_{lin}$ for each class of objects also translate
into digital tracking analyses with numbers of trial vectors that are not greatly
different: roughly 30,000 for MBAs, and a range of about 10,000 -- 200,000 for
both NEOs and KBOs depending on the survey parameters. The computational requirements
are greatest for NEOs because a single survey should include many, relatively
short digital tracking integrations --- but even for NEOs, the required processing
appears to be within the capabilities of modern clusters or supercomputers.

To give concrete examples,
digital tracking can be used to target
fast moving NEOs with a 1-hour integration composed of 120 individual
30-second exposures\footnote{Strictly speaking this would only work with
a next-generation imager having very short readtime.  With a 20-second
readtime (e.g. DECam), one could obtain 72 individual 30-second exposures
in an hour, and the digital tracking advantage would still be $\sqrt{72} = 8.5$,
close to a factor of 10.}; MBAs using a 6-hour integration composed of 180
two-minute exposures; or KBOs near opposition using a 36-hour integration composed of
216 ten-minute exposures acquired over six nights. The computational
requirement is the same to within a factor of a few for each of these integrations, and
would require less than a day of supercomputer time. Each integration will detect
objects more than ten times fainter than conventional observations
using the same telescope and SNR threshold.  Although conventional surveys can typically adopt a lower SNR
threshold (e.g. 5$\sigma$ for \citet{Denneau2013} versus 7$\sigma$ for our
own observations; see Section \ref{sec:comp}), this reduces the advantage of
a digital tracking survey only slightly.  For example, an object detected at 7$\sigma$ on a
stack of 180 images is $\sqrt{180} = 13.4$ times fainter than an equal-significance
detection on a single image, and remains 9.6 times fainter even than a 5$\sigma$ detection
on a single image. 

Digital tracking can also increase the sensitivity of surveys aimed at retiring terrestrial
impact risk by discovery and prompt followup of previously unknown NEOs. For such
surveys, the need for rapid data processing restricts the length of digital tracking
integrations: e.g., only 15-40 images would be stacked per integration (Section \ref{sec:risk}).
This reduces the likely sensitivity increase from digital tracking to factors of four to six,
rather than ten to fifteen for statistical surveys stacking over a hundred images.
Even with these limitations, digital tracking will still bring numerous previously
undetectable asteroids into range.

\section{Review of Past Work} \label{sec:overview}

\subsection{Kuiper Belt Objects and Outer Planet Moons} 

\citet{Tyson1992} are to be credited with the first application of digital tracking to Kuiper Belt searches, even though they did not detect any objects. There followed several other results in which a few KBOs were detected in deep, small-field digital tracking searches using large telescopes. \citet{Gladman1998} used the CFHT (Mauna Kea) and the 5m Hale Telescope (Palomar) to search 0.025 to 0.13 deg$^2$ per night for ten nights, finding 5 KBOs in all; \citet{Luu1998} used the 10m Keck Telescope (Mauna Kea) to search 0.01 deg$^2$ per night for three nights and found one object; and \citet{Chiang1999} used Keck to search 0.01 deg$^2$ over a single night and found two objects.  All of these surveys used the linear, constant-velocity approximation for the motions of their objects, and searched between 20 and 100 distinct trial stacks.  All of them also parameterized the motion space to be searched by the total angular velocity and its position angle, rather than by the eastward and northward components of the velocity vector. The former is appropriate in the context of the Kuiper Belt surveys, but the latter is preferable for the larger areas of motion space that must be searched in order to target asteroids.

As larger-format CCD imagers became available on large telescopes in the early 2000's, digital tracking surveys targeting KBOs put them to use.  \citet{Allen2001} used the 4-meter Blanco telescope to carry out the first (and, thus far, the only) survey to attempt digital tracking integrations spanning more than one night.  Because they also targeted three fields per night, their two-night integrations did not attain greater sensitivity than would a single night's observations targeting only one field; nevertheless their success proves the feasibility of multi-night integrations. Their digital tracking analysis used thousands of trial stacks, in order to search their angular motion phase space with sufficiently fine sampling to register the images of moving objects across a two-night span. Covering about 0.5 deg$^2$ per pair of nights over three different pairs, they detected 24 new objects. Other observers confined their digital tracking integrations to one night, but nevertheless obtained excellent results using the new large-format imagers.  \citet{Gladman2001} used the CFHT and the 8-meter VLT for two nights each, but analyzed the data from each night independently, detecting several new KBOs including some beyond 48 AU. \citet{Fraser2008} used the CFHT and the 4-meter Blanco Telescope to cover 0.32--0.85 deg$^2$ per night over a total of nine nights.  Using only 25 digital tracking trial stacks for each night's data, they covered a total of 3 deg$^2$ and discovered 70 new KBOs. \citet{Fraser2009} used the 8-meter Subaru Telescope (Mauna Kea) with its new Suprime-cam imager for a digital tracking search of 0.33 deg$^2$ of sky over two nights, finding 36 new KBOs. \citet{Fuentes2009} used archival data from the same instrument, combined with a digital-tracking analysis probing 736 trial stacks, to find 20 KBOs as faint as $R=26.8$. \citet{Kavelaars2004} and \citet{Holman2004} used digital tracking observations with the CFHT and the Blanco Telescope to find new irregular satellites of Uranus and Neptune, respectively, with each search covering of order 1 deg$^2$.

All of the above digital tracking surveys used the approximation of linear motion at a constant velocity. Except for \citet{Allen2001}, none required more than a few hundred digital tracking trial stacks to search the appropriate region in motion space.  By contrast, \citet{Bernstein2004}, building on an earlier attempt by \citet{Cochran1995}, used the Hubble Space Telescope to search a multidimensional angular motion phase space requiring about nine million digital tracking trial stacks per field. Covering a total area of 0.02 deg$^2$ with an effective integration time of 10.6 hr, they detected three new KBOs, including one at magnitude $m_{\mathrm{606W}}=28.38$ which remains the faintest Solar System object ever accurately measured.

\subsection{Asteroids} 

None of the above digital tracking observations targeted objects closer than the orbit of Uranus. Except for \citet{Kavelaars2004} and \citet{Holman2004}, all of them targeted KBOs. Indeed, until 2014 there was almost no attempt to apply digital tracking to asteroids. The only exception we are aware of is the work \citet{Gural2005}, who developed a mathematically sophisticated matched-filter technique for asteroid detection. They employed it in a manner equivalent to digital tracking, and demonstrated that they could detect previously-missed faint objects in existing data from the Spacewatch survey. As these data were optimized for conventional asteroid search methods, only 3--5 images of each field were available. The modest sensitivity increase possible in this context, and the differing optimal methodology for realizing it, renders the \citet{Gural2005} technique distinct from and complementary to digital tracking as discussed herein, even though it is closely analogous conceptually. Consistent with their own terminology, we shall not henceforth refer to the \citet{Gural2005} method as digital tracking, reserving the term for analyses that aim for large increases in sensitivity by stacking tens to hundreds of images of each field.

We are aware of only one published detection of a previously unknown asteroid using digital tracking as defined above. \citet{Zhai2014} achieved this detection using the 5m Hale telescope with the CHIMERA instrument described by \citet{Shao2014}. CHIMERA uses new EMCCDs that deliver extremely fast read times while maintaining low read noise.  The field of view at present is limited to 0.002 deg$^2$, but the instrument enables efficient observations with only a one-second exposure time.  This allows digital tracking observations targeting NEOs passing very close to Earth, with motions of several degrees per day.  In fact, while the single-frame exposure times for other digital tracking surveys are two minutes or longer, \citet{Zhai2014} stacked 30 one-second exposures for digital tracking stacks with an effective \textit{integration} of only 30 seconds. Despite this extremely short integration time, they detected a previously unknown 23rd magnitude asteroid moving at $6.32 \degree$/day (948 arcsec/hr). Due to the instrument's small field, which can only be partly compensated by the large number of short digital tracking integrations that can be carried out each night, this object is the only previously unknown asteroid to be detected with CHIMERA so far. As the maximum useful exposure ($\tau_M$) for such an object is less than four seconds, it would have been extremely difficult to detect with any other instrument.

The methodology of \citet{Shao2014} and \citet{Zhai2014} is highly complementary to our own, which uses large format CCD imagers to detect asteroids with larger geocentric distances and slower angular motions. While we have not targeted NEOs thus far, our methodology has the potential to detect them with good sensitivity at distances as small as 0.1 AU in fields of view hundreds of times larger than that of CHIMERA (Section \ref{sec:points}). For very small asteroids passing extremely close to Earth, CHIMERA and proposed next-generation instruments with larger fields \citep{Shao2014} will remain uniquely capable. If pushed to even shorter individual exposures, digital tracking using similar instruments would also detect small pieces of anthropogenic space debris in low Earth orbit.

\subsection{Past Work and the Potential of Digital Tracking} 

We are now in a position to compare the actual digital tracking surveys that have been carried out with our careful consideration of the limits of digital tracking in Sections \ref{sec:skymotion}--\ref{sec:points}. We find that although highly successful surveys have obtained scientifically valuable results on KBOs, the full potential of digital tracking has not been exploited even for them: considerably longer and more sensitive integrations are possible than have ever been attempted from the ground. Furthermore, while \citet{Zhai2014} have obtained a remarkable result in their detection of an NEO closer and faster-moving even than we have considered targeting, large-format CCD imagers can probe larger volumes of space and larger statistical samples of NEOs. No previous digital tracking survey has targeted either NEOs or MBAs with large-format imagers, except the MBA-optimized observations we report herein. For all three classes of objects, digital tracking holds yet-to-be exploited potential for extremely sensitive surveys.

\section{Our Survey of MBAs: Observations and Data Reduction} \label{sec:Data}

\subsection{Observations}
Our observations were obtained using the WIYN 0.9m telescope\footnote{The WIYN Observatory is a joint facility of the University of Wisconsin-Madison, Indiana University, the National Optical Astronomy Observatory and the University of Missouri.} on Kitt Peak, and the MOSAIC CCD imager.  MOSAIC is a 65 megapixel, 8-detector imager originally developed for (and mostly used on) the 4m Mayall telescope.  It has excellent cosmetic, QE, and noise characteristics, and a good readout time of 22 seconds. The unusual opportunity to use such an instrument on a relatively small telescope was the result of an agreement between the WIYN consortium and NOAO. On the WIYN 0.9m, the MOSAIC imager delivers a 59x59 arcminute field of view with 0.43 arcsecond pixels.

Since this was to be the first digital tracking survey for asteroids, we targeted MBAs rather than the less abundant and more challenging NEO population. We optimized our observations for MBAs both in terms of the integration lengths (all night) and the individual exposure times (2 minutes: i.e., $\sim \tau_M$ for MBAs). Although in the current work we analyze only MBA detections based on WIYN 0.9m data, we note that NEOs can be detected by digital tracking analysis even of MBA-optimized data sets, and we present such detections from a DECam data set in a forthcoming work (Heinze et al, in prep). 

To detect the maximum possible number of faint asteroids, we centered our WIYN 0.9m observations on the ecliptic, at a point directly opposite the Sun. Such a field would be highest in the sky and allow the longest possible integration times in December (for the Northern Hemisphere), when the coordinates would be near RA 06:00 and DEC +23:15. However, near RA 06:00 the ecliptic is crossing the Milky Way, producing extremely rich starfields that create challenges for asteroid detection. Although we believe that the star subtraction methods we are developing (Section \ref{sec:starsub}) will allow digital tracking to perform well even in rich starfields, we desired sparse starfields for our first survey. Thus, we chose to observe in the spring even though antisolar fields are then at more southerly declinations.

On the nights of April 19 and April 20, we observed a single ecliptic field all night, accumulating roughly five hours' worth of integration.  Moonlight was bright enough to reduce our sensitivity appreciably (full Moon was on April 24), but the weather was good and the seeing averaged 1.6--1.7 arcseconds through the majority of each night. On April 19, we acquired 150 two-minute R-band exposures of a field centered near RA 13:56:13, DEC -11:53:22; and on the following night we acquired 158 exposures of a field centered near 13:55:23, -11:49:14. Both fields are very near the antisolar point on their respective dates. The offset between them tracks the mean sky motion of main belt asteroids, determined based on an average of known asteroids near these fields on the dates of our observations.  As the motion of such asteroids over 24 hours was much less than the field of the MOSAIC imager, the two fields overlap heavily; nevertheless the offsetting is desirable to avoid unnecessary loss of asteroids near the field edges from one night to the next.

We paid careful attention to dithering the images, both to fill the gaps between the detectors in the MOSAIC imager and to allow the construction of star-free night sky flats from a median stack of the science images. Dithering is also desirable because it tends to convert systematic effects into random effects, and greatly reduces the chance that detector artifacts will produce spurious detections in stacked images. We performed a dither offset every two images, using a quasi-random pattern centered on the coordinates given above and spatially constrained so that all pointings lay within 5 arcminutes of the nominal position.  The dithering was quasi-random in that it consisted of many sets of regular dither patterns (e.g. hollow squares and linear dithers along various vectors), but the spacing and angles describing these patterns were changed to avoid redundant pointings and thoroughly sample the spatially constrained dither region.

\subsection{Image Processing}

\subsubsection{Basic Processing}
We begin our image processing by subtracting the mean in an overscan region for each detector, interpolating across the few cosmetic defects in the MOSAIC detectors, and correcting for electronic crosstalk. Crosstalk in the MOSAIC detector is relatively simple: bright sources in one half of each detector produce spurious mirror-images of themselves at greatly reduced brightness in the other half. It is easy to remove these spurious images by subtracting from each detector half-image a reflection of the other half multiplied by the appropriate crosstalk coefficient. We determine the crosstalk coefficients for each detector empirically from our data, finding that they range from 0.0016 to 0.0021 and exhibit no significant variations over time.

Following the crosstalk correction, we use the Laplacian edge detection algorithm of \citet{lacos} to remove cosmic rays\footnote{The order of operations is important here:  if the \citet{lacos} cosmic ray removal is applied first, it will trigger on the edges of the crosstalk artifacts from saturated stars and partially remove them, after which the crosstalk correction will introduce artifacts by attempting to subtract artifacts that have already been partially removed.}.  Following these operations, the individual images from the eight MOSAIC detectors are tiled into single-extension FITS images with pixel dimensions 8,192$\, \times \,$8,192. This allows us to perform dark subtraction and flatfielding using simple routines designed for single-detector cameras --- in particular, it means that a simple normalized flat automatically corrects for variations in sensitivity between the detectors.  The flatfield we use is a clipped median-stack of dark-subtracted science images.  Due to the thorough dithering described above, the stars vanish from this stack. After correction using this flatfield, our images have fairly uniform sky backgrounds. We find that removing the sky background by simply subtracting a constant value equal to the clipped median over all pixels in the image is sufficient for our purposes.

\subsubsection{Astrometric Registration}
At this point, our tiled images are fully corrected and ready for astrometric registration.  We begin by separating them again into the tiles corresponding to each of the eight detectors of the MOSAIC imager.  Based on a manually selected reference star, for each tile we identify stars in the UCAC4 astrometric catalog and use them to construct an astrometric solution mapping pixel coordinates to celestial coordinates.  We use a third-order polynomial mapping with cross terms, yielding ten degrees of freedom. To reduce the danger of roundoff errors in fitting, we measure both pixel and celestial coordinates relative to the center of the image, and we apply scaling factors so that the absolute values of all coordinates are near 1.0 at the image edges. We iteratively reject stars from the fit until the worst remaining outlier deviates by less than 0.3 arcsec. Since typically 50-80 stars per detector survive this clipping, the fits remain well constrained.

\begin{figure} 
\plottwo{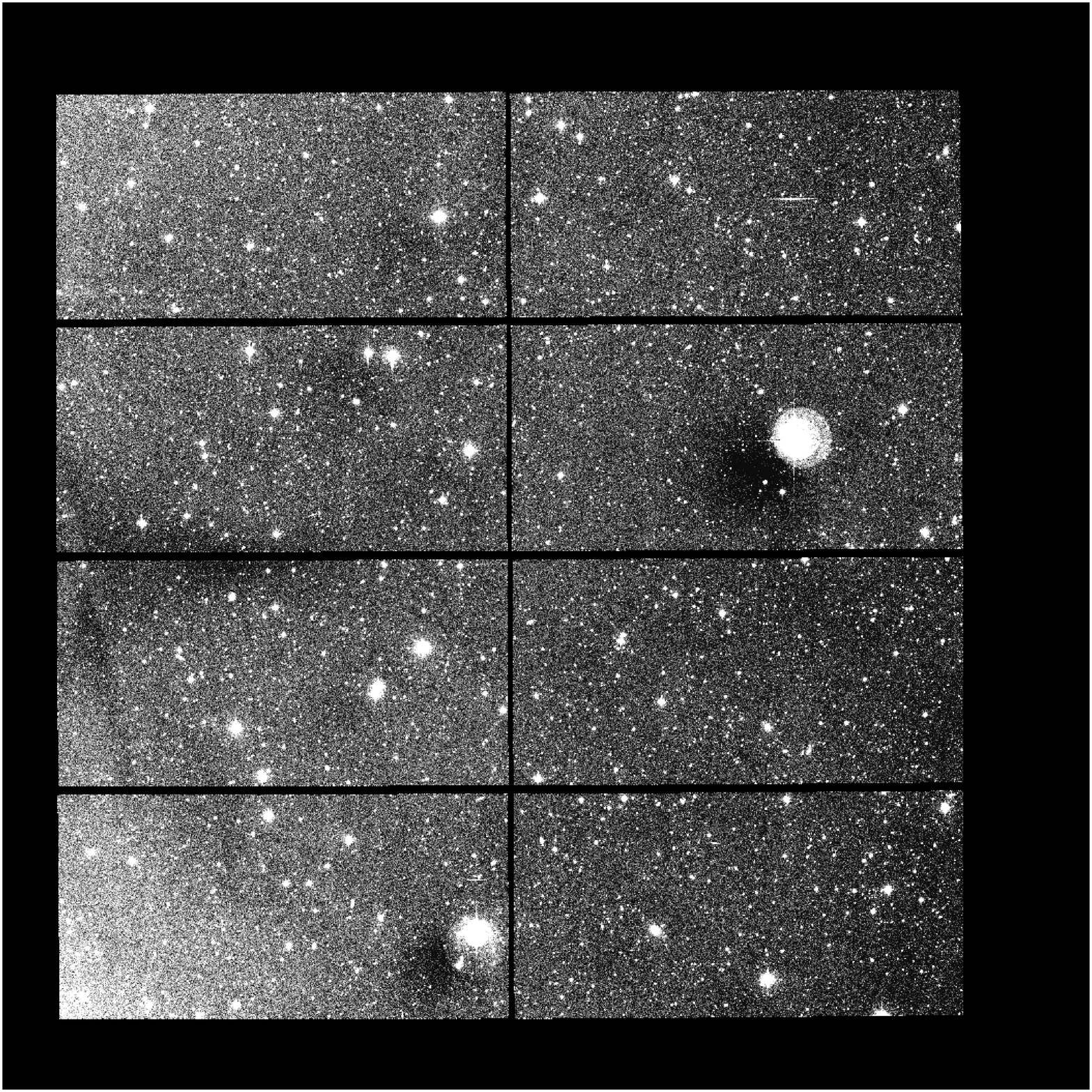}{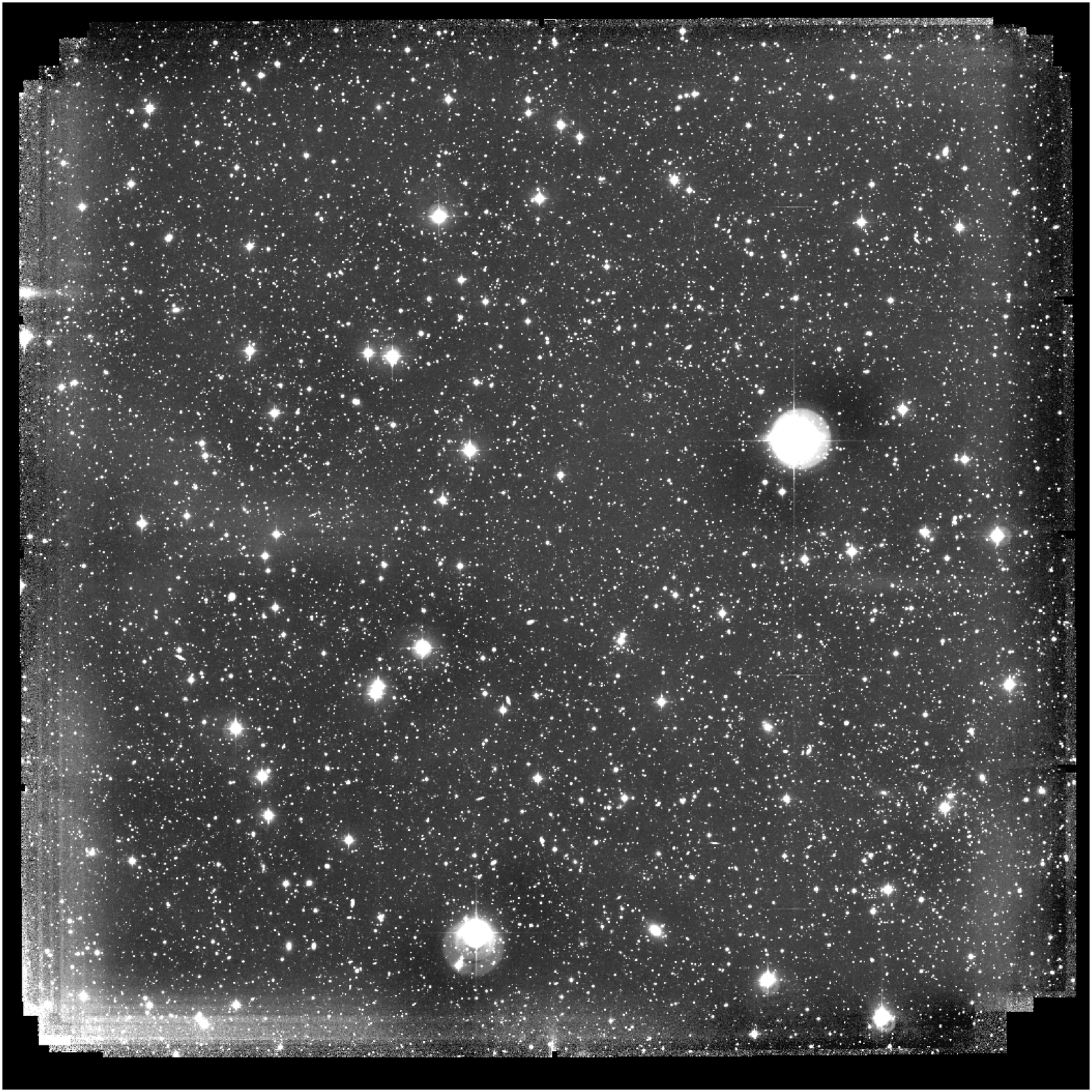}
\caption{MOSAIC images resampled onto a consistent astrometric grid. Both panels
show a 70$\times$70 arcminute region corresponding to 10,500$\, \times \,$10,500 resampled pixels; excess
zero-padding in the original 12,000$\, \times \,$12,000 resampled image has been trimmed away.
\textbf{Left:} A single 2-minute exposure image from April 20.  Gradients and other
diffuse features in the sky background are not real, but such artifacts are irrelevant
to our analysis. The eight individual detectors are clearly seen. \textbf{Right:} The master image
made by stacking all 158 April 20 images like that at left.  The gaps are filled and
the region of useful data somewhat expanded due to our dithering strategy.
\label{fig:ditherdat}}
\end{figure}

Having obtained an astrometric fit for each detector, we resample each image onto a consistent astrometric grid using bilinear interpolation. We define this grid to have a constant pixel size $s_{pix}$ of exactly 0.4 arcsec, such that the $x$ and $y$ pixel coordinates are given using the simple mapping

\begin{equation} \label{eq:mapping}
\begin{array}{lcl}
x &=& x_0 - (\alpha - \alpha_0) \cos(\delta)/s_{pix} \\
y &=& y_0 + (\delta - \delta_0)/s_{pix} \\
\end{array}
\end{equation}

\noindent where $\alpha$ is the RA; $\delta$ is the DEC; and $x_0,y_0$ and $\alpha_0,\delta_0$ both refer to the center of the new, resampled image (which may be freely chosen by the user); and the difference in sign between the equations is needed to avoid mirror imaging and to produce a final image with east left and north up.  The output image, which contains the data from all eight individual detectors resampled onto the single self-consistent grid defined by Equation \ref{eq:mapping}, measures 12,000$\, \times \,$12,000 pixels.  This includes the gaps between the detectors, as well as generous zero-padding that we include around the outside of the mosaic to prevent any data from dithered images from being lost off the edges of the array. Figure \ref{fig:ditherdat} illustrates our resampled images, with excess zero-padding trimmed away.

\subsubsection{Preservation of Geometric Linearity under Resampling}

The astrometric grid on which we have resampled our images constitutes a very simple map projection of the spherical sky onto a two-dimensional plane. Our digital tracking algorithm makes the approximation that the objects being sought will move with constant angular velocity in a straight line. In Section \ref{sec:skymotion}, we explored the range of validity of this assumption in terms of the sky motion of the objects, and found it sufficient to allow extremely long digital tracking integrations.  Now, however, we must consider another question: whether Equation \ref{eq:mapping} projects linear, constant-velocity sky motion onto linear, constant-velocity motion across the pixel grid with sufficient accuracy. 

In evaluating a map projection for digital tracking images, only two types of distortion are relevant: spatial variations in the pixel scale, and failure to map linear (i.e. Great Circle) trajectories onto straight lines on the pixel grid.  No projection is simultaneously free from both types of distortion. We have chosen the projection described by Equation \ref{eq:mapping} because it is mathematically simple and is free from distortion of the first type: all of the pixels have exactly the same angular size. However, it does suffer from distortion of the second type. While lines of constant RA are Great Circles and lines of constant DEC (other than the equator) are not, our projection renders lines of constant DEC as straight and lines of constant RA as curved except for the central one that vertically bisects the image. In the following, we quantify the importance of this distortion for digital tracking analyses.

The relevant value is the maximum extent by which the projection of a track that is a segment of a Great Circle deviates from a straight line. We have calculated this deviation for a wide range of cases in order to probe the limits of our map projection, fixing the track length to be 20 arcminutes because the maximum digital tracking integration lengths described in Section \ref{sec:skymotion} correspond to tracks of this length or shorter for all classes of objects. The deviation is strongly dependent on the celestial position angle of the track, being in general much worse for east-west motion than for north-south motion, due to the characteristic of Equation \ref{eq:mapping} that lines of constant DEC are rendered as straight lines even though they are not Great Circles. Great Circle tracks with a locally east-west orientation therefore show deviations at all nonzero declinations even if they pass through the center of the field. By contrast, north-south trajectories follow RA lines, which Equation \ref{eq:mapping} renders correctly at the center of the field, so such tracks have nonzero deviations only when they are offset in RA from the field center.

Objects can exhibit east-west trajectories in any digital tracking observation, so this worst-case scenario is relevant. The maximum deviation for a 20 arcminute Great Circle track is zero on the celestial equator, but reaches 0.5 arcsec at $40 \degree$ DEC and 1.0 arcsec at $60 \degree$.  The deviation of north-south trajectories is independent of DEC, is always zero at the center of the field, and rises only to 0.05 arcsec for an RA offset of $5 \degree$ from field center. Thus, the map projection given by Equation \ref{eq:mapping} is adequate for digital tracking applications up to $40$--$60 \degree$ DEC with fields spanning more than $10 \degree$ in RA --- far larger than the field of view of any major telescope. For our own observations using a one square-degree imager at $-12 \degree$ DEC, targeting asteroids whose track lengths are less than 6 arcminutes, track deviations from the Equation \ref{eq:mapping} projection are entirely negligible, of order 0.01 arcsec. 

The projection given in Equation \ref{eq:mapping} is not appropriate for digital tracking observations at declinations greater than $40$--$60 \degree$. Alternatives may easily be envisioned: for example, one can perform the astrometric resampling in the local tangent plane and then transform the positions and motions of detected asteroids back to the celestial coordinate system once the digital tracking analysis is complete. This is equivalent to using Equation \ref{eq:mapping} on the celestial equator: even with a $10 \degree$ field, the maximum distortion is only 0.05 arcsec. Mapping distortion therefore does not necessitate any reduction in the maximum durations for digital tracking observations that were calculated in Section \ref{sec:skymotion}.

\subsubsection{Subtraction of Stationary Objects} \label{sec:starsub}
While not all previous digital tracking searches subtracted stationary objects prior to the digital tracking analysis, our large-scale implementation renders such subtraction essential.  This is because we need to be able to avoid unmanageable numbers of false positives while at the same time using a fast and simple automated algorithm to search each of our thousands of trial stacks --- and simple algorithms are easily confused by the noisy streaks that unsubtracted stars and galaxies leave behind, even in clipped-median stacks.

\citet{Alard1998} and many others have developed sophisticated techniques for subtracting constant stars and galaxies from a series of images in order to find variable objects even in highly confused fields. To measure variable stars in a consistent way, these methods require the use of a consistent reference image for subtraction throughout the data set. Normalized convolution kernels are then identified that will blur the reference image to match each successive image from which it is to be subtracted. Our case is different in two ways. First, we are not trying to measure stationary objects, only to remove them as completely as possible: we do not have to use a consistent reference image. Second, we place a very high priority on achieving the subtraction with the least possible increase of sky background noise.

Rather than the convolution strategy of \citet{Alard1998}, we choose to model each image in our data set as a linear combination of other images taken at a large enough temporal separation that moving objects in our target populations (the MBAs) cannot self-subtract. We find that conventional least-squares methods for determining the optimal linear combination severely amplify the noise in some cases by producing large positive and negative coefficients with similar absolute values. Hence, we use a downhill simplex method from \citet{nrc} to find the best least-squares match with coefficients restricted to positive values. To further reduce noise and speed processing, we divide the set of images from each night into several (typically seven) contiguous subsets and create a clipped median stack of each subset, producing a smaller number of cleaner and lower-noise star images. The downhill simplex analysis then matches each individual image using a linear combination of these seven low-noise star images. Since the PSF can vary across an image, we perform the downhill simplex matching individually within sixteen regions defined by an approximately rectangular Voronoi tessellation across our image. In order to avoid fitting irrelevant sky noise and accurately match the PSFs of stars, pixels with values below a threshold on the master stack (shown in the right panel of Figure \ref{fig:ditherdat}) are excluded from the fit. As a final step, a constant value is subtracted to set the median sky value to 0.0 in each Voronoi region.

\begin{figure} 
\includegraphics[scale=0.32]{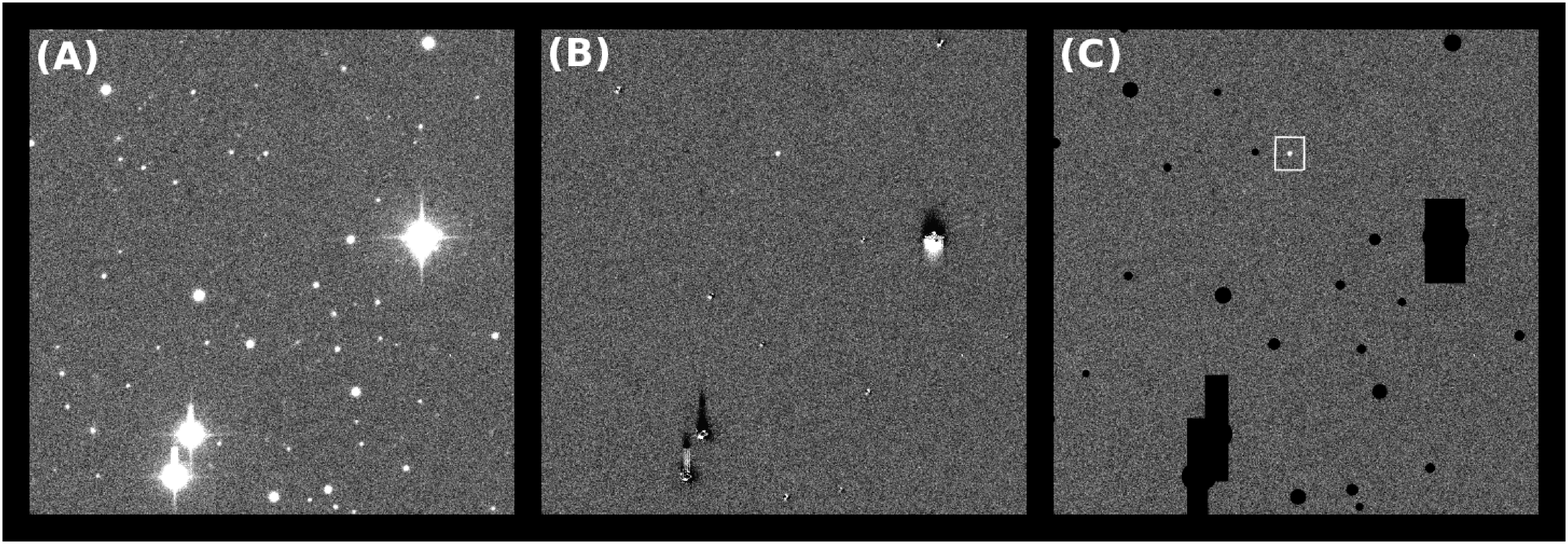}
\caption{Subtraction and masking of stationary sources illustrated by a small region
trimmed from a single frame from our April 20, 2013 data set. The unsubtracted object
boxed in the rightmost panel is an asteroid bright enough to be detected without digital
tracking. (A) Original resampled image. (B) The same image after subtraction.
Artifacts, probably electronic in origin, can be seen emanating from the brightest stars.
(C) The same image after the final masking step, ready for use in our digital tracking
analysis. 
\label{fig:subtraction}}
\end{figure}

This subtraction eliminates faint stars and galaxies completely, but noisy residuals remain near the cores of bright objects, as well as inconstant artifacts that are probably electronic in origin emanating from saturated stellar images. We mask these residuals aggressively based on pixel brightnesses in the master star image, adding rectangular regions by hand for the worst saturated objects.  Figure \ref{fig:subtraction} illustrates our results.

While the image subtraction methodology described above was entirely satisfactory for our April 2013 data, we have continued developing image subtraction for future, more challenging applications. Full discussion of this development is beyond the scope of the current work, but we will make two important observations.  First, our Voronoi tessellation turns out to be an irrelevant frill: a simple rectangular grid works fine.  Second, a hybrid method including the kernel convolution developed by \citet{Alard1998} produces much better results in crowded fields at low galactic latitude. Although this method greatly increases the computer time for image subtraction, it remains small compared to that required for the digital tracking analysis itself.

\section{Digital Tracking Analysis}

\subsection{Basic Parameters} \label{sec:digipars}
As discussed in Section \ref{sec:skymotion}, digital tracking integrations targeting main belt asteroids can last 8--9 hr within a single night, but cannot span multiple nights under the approximation of linear, constant-velocity motion.  Thus we analyze our data from April 19 and April 20 separately.  The two data sets consist of 150 and 158 images, respectively, each a 2-minute exposure in the $R$ band.  

We excluded some images on each night that exhibited elevated sky backgrounds (due to twilight) or had unusually bad seeing, in order to avoid wasting compute time stacking poor-quality images that would add little or nothing to our final sensitivity. On April 19, we excluded the final 24 images of our data sequence on account of very bad seeing (averaging 2.8 arcsec vs. 1.6 arcsec for the 126 remaining images). On April 20, we excluded 13 images from the beginning of the data sequence due to elevated background levels (averaging 1,600 ADU vs. 1,000 ADU for the remaining images) and 15 images from near the end of the sequence due to bad seeing (averaging 2.5 arcsec vs. 1.7 arcsec for the 130 remaining images). Thus we arrived at final data sets consisting of 126 images for April 19 and 130 for April 20. 

The time spanned by these data sets is 5.76 hr in each case. Equation \ref{eq:blur} indicates an optimal grid spacing of $\Delta_m = 0.25$ arcsec/hr for the digital tracking search, and we conservatively adopt $\Delta_m = 0.20$ arcsec/hr. We search a rectangular region in angular motion phase space extending from -50 to -20 arcsec/hr eastward and -7 to +30 arcsec/hr northward, which spans the range of plausible motion vectors for MBAs (Figure \ref{fig:drift}).  In all, we probe 28,086 trial vectors.

We perform our digital tracking analysis using a desktop workstation having 64 gigabytes of memory.  This memory is insufficient to simultaneously load and manipulate 130 full-size resampled images, so we divide each image into four overlapping quadrants of 6,000 $\times$ 6,000 pixels. We are able to make the quadrants overlap by discarding some of the excess zero-padding in our original 12,000 $\times$ 12,000 pixel resampled images. The overlapping is necessary in order to maintain full sensitivity for asteroids that cross a quadrant boundary.

Thus, for each data set we carry out four digital tracking runs, each processing a single quadrant. Since each quadrant is processed through 28,086 trial vectors, our search over four quadrants probes 112,344 trial stacks with a size of 6,000 $\times$ 6,000 pixels for each of the two data sets. The total number of vector-pixels (number of images $\times$ pixels per image $\times$ number of trial stacks; see Section \ref{sec:mbatrial}) probed in our two data sets of 126 and 130 images is $1.035\times10^{15}$. We use nearest-neighbor interpolation for our image shifts and a median with 5$\sigma$~clipping for our stacks. Nearest-neighbor interpolation speeds processing relative to bilinear interpolation, and is not expected to reduce sensitivity since our images are well sampled. Our full digital tracking analysis requires a total of about 50 days of runtime for our multi-core desktop, which processes the data at a rate of $8.1 \times 10^{11}$ vector-pixels per hour.

\subsection{Automated Detection of Asteroids}

\subsubsection{Choosing a Detection Threshold} \label{sec:autothresh}
Taking into account the zero-padding and overlapping of the quadrants, our search of 224,688 trial stacks (that is, 112,344 stacks per night for two nights of data) spans about $5.6\times10^{12}$ non-overlapping non-zero pixels in the final stacks. These correspond to $6.2\times10^{11}$ square, 3 $\times$ 3 pixel boxes that each sample the noise over a region equal to the size of an asteroid's PSF. If all of these noise realizations were independent and the noise were Gaussian, detection thresholds of 8$\sigma$, 7$\sigma$, or 6$\sigma$ would result the expectation of $4 \times 10^{-4}$, 0.8, or 600 false positives, respectively, for our entire survey. Our sampling of angular motion phase space is fine enough that adjacent trial stacks are not completely independent, which produces a modest reduction (no more than a factor of two) in the expected false positive numbers. While false positives due to non-Gaussian effects such as edge noise, star subtraction residuals, or cosmic rays would be rejected by the manual checking described below (Section \ref{sec:mancheck}), pure Gaussian outliers might not exhibit any signatures of problematic data and hence could pass all the tests. We wish to detect as many genuine asteroids as possible while not reporting any spurious detections as real objects, and for simplicity we prefer to choose an integer threshold. Thus, we adopt a 7$\sigma$ threshold for our automated detection: it is the integer value that yields the greatest sensitivity while still producing an expected number of false positives less than 1.0. We will conclude below (Section \ref{sec:falsepos}) that none of the asteroids we finally confirmed is a false positive.

\subsubsection{The Automated Detection Algorithm}
Saving trial stacks for later analysis would require prohibitive amounts of hard drive space, so our digital tracking code searches and then discards each stack, ultimately outputting not a set of images but a detection log.  To streamline processing, we use a simple and fast algorithm for source detection. First, the trial stack is smoothed with a square boxcar approximating the size of the PSF: in this case, 3$\times$3 pixels or 1.2$\times$1.2 arcseconds. A new image that maps the standard deviation of this smoothed image is then constructed with a resolution of 3 pixels.  We obtain the value for each pixel in this map by calculating the standard deviation in a square annulus around the corresponding point on the smoothed image: the annulus has inner dimension 15 pixels and outer dimension 27 pixels. Within this annulus, only every ninth pixel (every third pixel in each dimension) is used in calculating the standard deviation, to preserve independence. The smoothed image is then scanned for pixels whose brightness indicates a detection above 7$\sigma$ based on the standard deviation map. If such a pixel is also the brightest one within a `redundancy radius' of 5 pixels, it is written to the log as a possible asteroid detection. 

\subsubsection{Duplicate Detections of Real Asteroids}
Our simple and fast detection algorithm results in numerous duplicate detections of real asteroids, and these are the dominant type of `false positive' encountered in the automated detection logs. The brightest objects produce many hundreds of duplicate detections because even on trial stacks far from the correct motion vector, the asteroids' streaked images remain above the detection threshold. Rather than adopting a more sophisticated (and therefore slower) automated detection algorithm, we use several techniques to winnow down the thousands of automatically logged detections to a much smaller set consisting exclusively of real asteroids.

The first step is to gather duplicate detections into clusters and retain only the brightest detection at the center of each cluster. Our clustering code begins with the very highest-significance detection in the whole log, which corresponds to a bright real asteroid on an accurately motion-matched trial stack.  A radius is defined surrounding this asteroid, and all objects with pixel coordinates lying inside this radius are provisionally classified as duplicate detections. Finally, the standard deviation of motion rates among these provisional duplicates is calculated, and 3$\sigma$ outliers are re-classified as possibly real, fainter asteroids that passed near the much brighter object. The program then proceeds to the most significant remaining un-clustered detection, and makes it the center of a new cluster.  

The radii used for clustering are determined as follows. First, the cluster program calculates the mean density of detections in pixel space based on the entire log.  Next, the user specifies a threshold factor above this background density\footnote{We find the optimal threshold factor is near 15 for the current data set, but it could be widely different in other contexts.}.  When defining a cluster, the program expands or contracts the bounding radius until the density of detections within it reaches the specified threshold above the background density.  When the detection log is fully clustered, the program outputs the fraction of the entire 4-dimensional\footnote{The four dimensions are the ordinary x and y dimensions of the images plus the two dimensions of the angular velocity space.} volume of the digital tracking space that was ultimately included in a cluster.  This represents an estimate of the false-negative rate (FNR) due to distinct real asteroids being incorrectly clustered with brighter objects.  If the threshold factor above the background density is too high, clusters are too small and the manual effort required to weed out the resulting large number of duplicate detections for bright asteroids becomes excessive. If the factor is too low, clusters are too large and the FNR becomes excessive. We optimize the factor manually to obtain a manageably small number of duplicate detections without raising the FNR above 1\%. In some cases, a satisfactory clustering may be obtained while the estimated, cluster-induced FNR is still as low as 0.1\%. In others, in order to reduce the numerous duplicate detections we lower the threshold factor until the FNR rises to 1\%, but in no case do we allow the FNR to exceed 1\%. The final output of our clustering code is a greatly refined list of detections which are likely to be real, motion-matched asteroids.

\subsection{Manual Checking of Detections} \label{sec:mancheck}
Each detection output by our cluster code is specified by its motion rates and its pixel coordinates on a reference image from the middle of our data sequence. We have produced visualization software that uses this information to quickly reconstruct a small region of the appropriate trial stack, centered on the asteroid.  Using the output of the cluster code, this allows us to reconstruct --- in a few minutes --- images that show all the plausible asteroids detected in a digital tracking search that ran for weeks. We manually examine such images to confirm or reject each candidate asteroid. Since our visualization software uses bilinear interpolation for the image shifts, this test also allows us to check reported detections using a different interpolation scheme from the nearest-neighbor method adopted in the initial digital tracking search.

\subsubsection{Motion Rate Check Images} \label{sec:checkim}
\begin{figure} 
\includegraphics[scale=0.35]{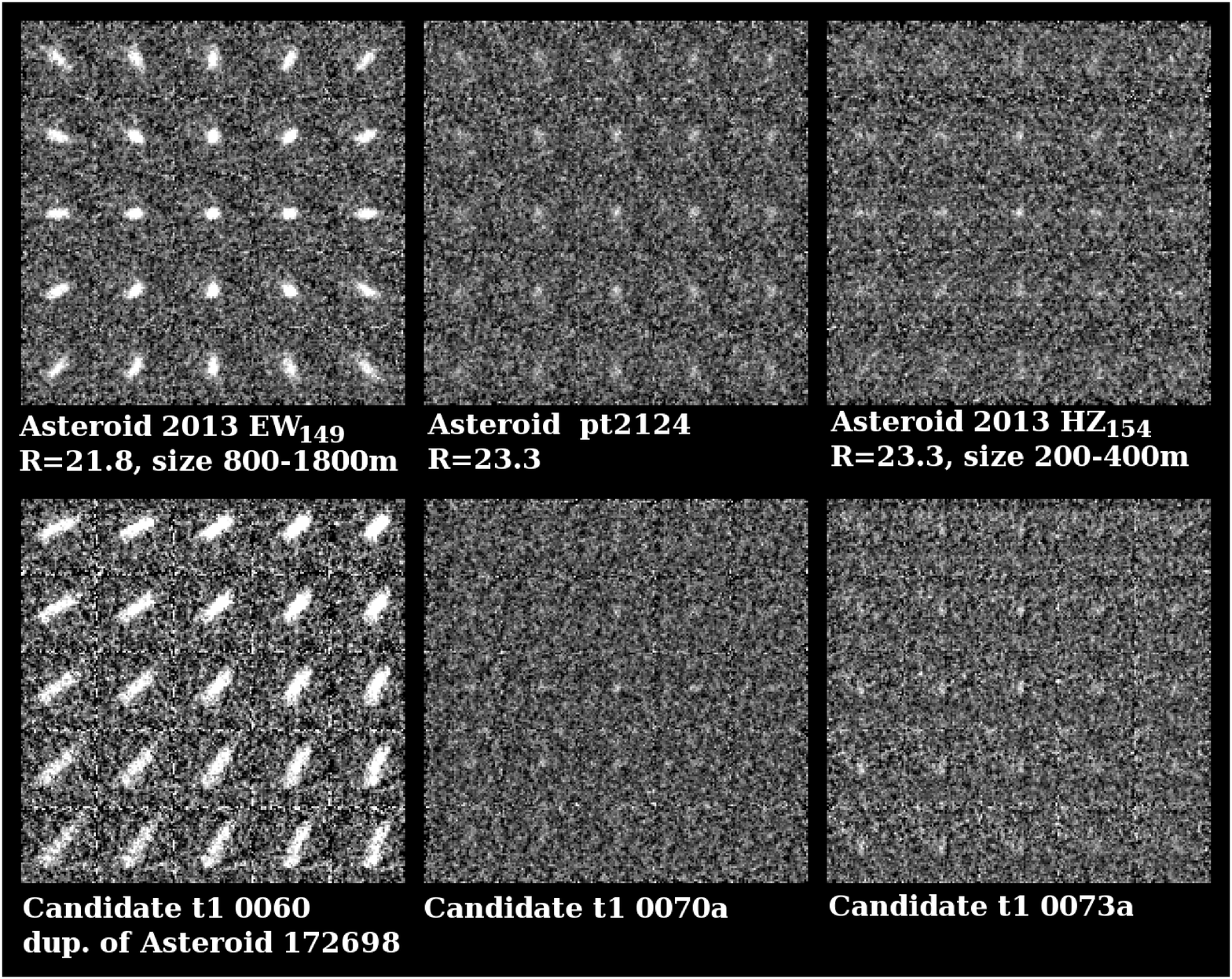}
\caption{Examples of check images used to manually verify automatically detected asteroids.
In each panel, the different tiles are offset from one another successively by 0.4 arcsec/hr
in angular velocity space, and the central one has the velocity of a particular automated detection. 
All objects in the top row are real, and only 2013 EW$_{149}$ was known prior to our analysis. Asteroid pt2124 was detected only on April 19 and thus lacks size and distance measurements, 
which require two nights. Objects in the bottom row are dubious or spurious, exemplifying categories
(2), (3), and (4) from Section \ref{sec:checkim}. Candidate t1 0060 is
a badly motion-matched duplicate detection of a known $R=20.4$ mag asteroid. Candidate t1 0070a may be real, but is very faint. Candidate t1 0073a does not fade uniformly toward every corner:
it may be a noise artifact.
\label{fig:check}}
\end{figure}

For our first test, we produce a small (e.g. 41$\times$41 pixel) stacked image centered on the asteroid under investigation using the motion vector corresponding to its logged detection, and several other identically-centered images using motion vectors corresponding to adjacent gridpoints in angular motion phase space. We tile these small images together, placing the nominally motion-matched tile in the center and arranging the others around it such that that if the detection is real, the streaked images in the mismatched tiles will point back toward the sharp image at the center\footnote{A similar way of checking digital tracking detections was independently developed by \citet{Fraser2008}; see their Figure 2.}. For convenience, we will refer to such tiled images as check images. Looking at a check image allows us immediately to determine if the asteroid is (1) a real, motion-matched detection; (2) a duplicate, velocity-mismatched detection of a bright asteroid that slipped through the clustering code; (3) an apparently real detection that is too faint to confidently classify; or (4) a completely spurious detection due to a noisy edge or other artifact in the trial stack. The top row of Figure \ref{fig:check} shows check images of three real asteroids, while the bottom row shows dubious or spurious detections: one example each of categories (2), (3), and (4).

Our program for making check images also performs a quadratic fit to the flux of the central source in each tile to determine the true flux-maximizing motion rates with greater accuracy than can be achieved by eye. The flux is measured within a small aperture of radius only 2.0 pixels (0.8 arcsec), so even small motion errors blur the image enough to reduce the measured flux. We iterate with the check image program as necessary, refining the motion rates to ensure that the optimal motion identified by the quadratic fit lies near the center of the fitting region. Finally, we pass all detections in categories (1) and (3) on to the second stage of manual verification.

\subsubsection{Verification in Independent Subsets of the Data: Pyramid Images} \label{sec:pyrim}
In the second round of verification, we again create multiple small images from digital tracking stacks, but instead of using different motion rates, we keep the rates fixed at the optimal values and use different subsets of the input images. We divide the full data set into first two, then three, then four, etc. consecutive independent subsets. For easy manual investigation, we arrange these images in a pyramid with the stack of all images by itself at the top; the two half-stacks below it; the three one-third stacks below that, etc.  We use five-layer pyramids for our analysis, but the bottom layer is needed only in special cases (see below). Figure \ref{fig:pyr} gives examples of four-layer pyramid images from our April 19 data. As this data set contains 126 images, the numbers of images per tile in the respective layers is 126, 63, 42, and 31 or 32. The corresponding integration times are 4.2, 2.1, 1.4, and 1.05 hours.

\begin{figure} 
\includegraphics[scale=0.26]{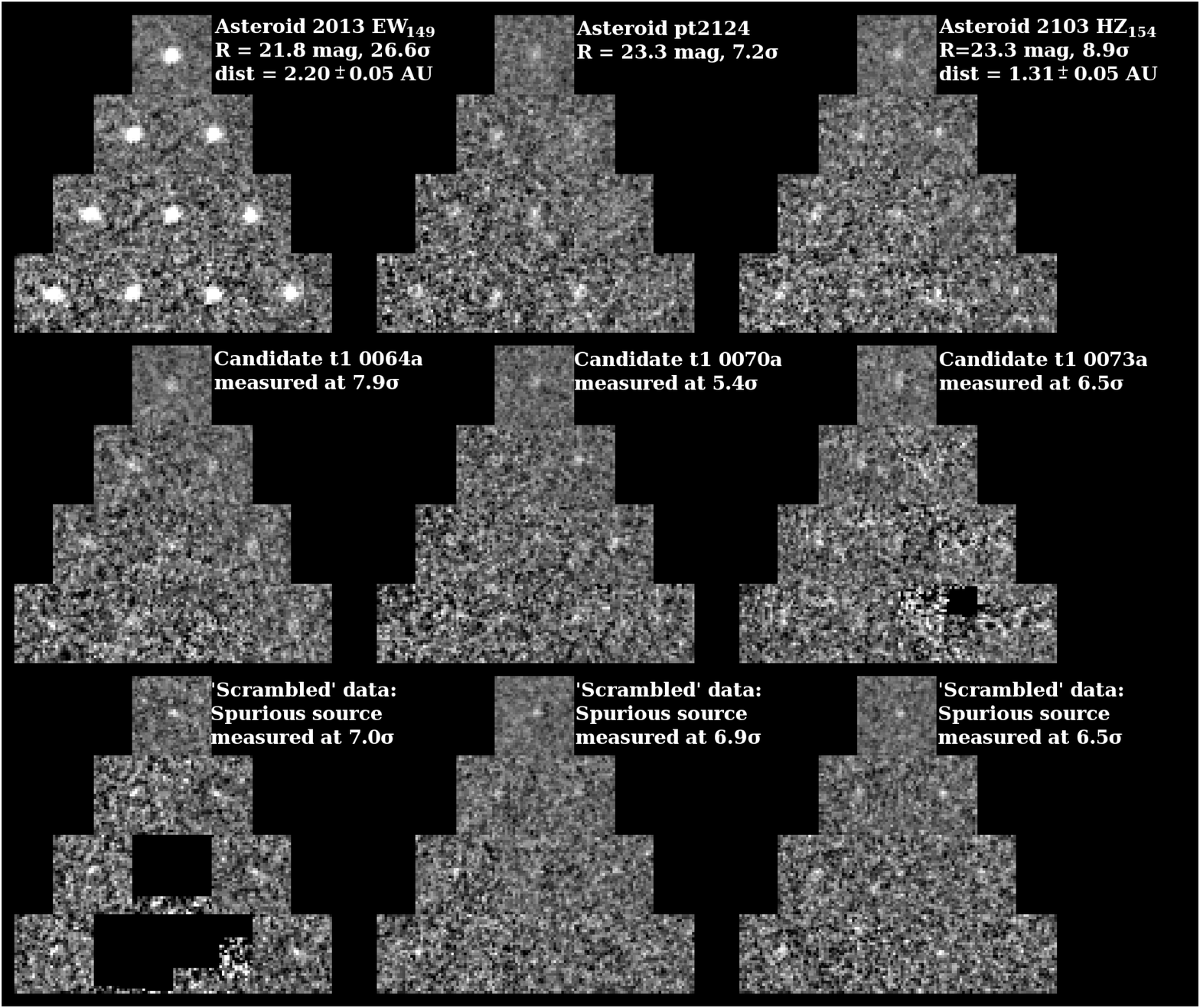}
\caption{Examples of pyramid verification images from our April 19 data. 
As described in the text, each layer of a pyramid represents a different division of our 126 images into equal independent subsets. Appearance in multiple tiles of the same layer confirms an object's reality. Objects in the top row are confirmed real, pt2124 and 2013 HZ$_{154}$ being new discoveries. Those in the middle row are not confirmed: t1 0064 is  probably real but is too faint for secure detection, 0070a is less plausible, and 0073a appears to be an artifact of noise from the masking of a bright star. We do not include unconfirmed objects in Table \ref{tab:asteroids}, nor have we reported them to the MPC. The objects in the bottom row are all spurious: they are the three most significant detections from our `scrambled' data set (Section \ref{sec:falsepos}). Significance values quoted in units of $\sigma$ are not from the raw detection logs but are re-calculated more accurately using master stacks produced by our visualization software.
\label{fig:pyr}}
\end{figure}


The purpose of the pyramid images is to test if a detection can be confirmed in multiple independent subsets of the data. It is difficult to conceive of an artifact that could imitate a real asteroid well enough to pass this test.  A spurious detection caused by a coincidence of cosmic rays, CCD defects, or star-subtraction residuals would reveal its spurious nature by vanishing from some tiles but getting brighter (and likely changing morphology) on others. Our thorough dithering, star subtraction, and cosmic ray removal renders such events vanishingly rare in our data. More common are spurious detections caused by locally noisy regions close to smoother data that cause our automated algorithm to underestimate the noise. Candidate t1 0073a in Figure \ref{fig:pyr} is an example: it appears similar to confirmed asteroid pt2124 in the full stack, but in the lower rows of the pyramid we see that it coincides with the masked region surrounding a bright star, and may be an artifact of the locally higher noise due to reduced data coverage at this location.

All other things being equal, a real object that is detected with with significance $s$ in the top tile of a pyramid image should appear at significance $s/\sqrt{2}$ in each of the half stacks, $s/\sqrt{3}$ in the one-third stacks, etc. This idealized situation frequently fails to be realized in practice. In extreme cases (e.g. objects that entered or left the field during our observations), only half the exposures may have supplied image data, but the asteroids can still be confirmed as real if they are bright enough. This is our reason for using five-layer pyramids in our analysis: the bottom layers can be necessary for verifying such objects even though four-layer images like those shown in Figure \ref{fig:pyr} are sufficient for most others. More commonly, rotation can cause asteroids to change brightness considerably on a timescale of hours, and changes in the seeing and sky background can change the sensitivity in different subsets of the data. Our confirmation criterion for pyramid images is therefore simply that the object must be consistently detected and readily apparent to the human investigator on at least two different tiles in a single layer of the pyramid. Provided this condition is satisfied, we do not penalize objects for not being readily visible in additional tiles in the same layer. It would be reasonable to apply such a penalty only if the implied change in brightness was implausibly large, which is a scenario we have not encountered. 


\section{Results for Detected Asteroids} \label{sec:results}

\subsection{Initial Matching to Known Objects} \label{sec:known1}
The criteria described in Section \ref{sec:pyrim} resulted in 199 confirmed asteroids in our April 19 data (Figure \ref{fig:tracks}), and 181 in our data from April 20. The slight reduction in sensitivity from one night to the next is likely due to the waxing of the moon. Even on April 19, the 10-day-old moon was bright enough to reduce the sensitivity of our observations significantly from their full dark-sky potential, and on April 20 the moon was brighter and closer to our field. 

We compare our final detection lists against ephemerides from the Minor Planet Center (MPC), matching objects within a 20 arcsecond radius and also checking for consistent sky motions. We find that on April 19, we detected 47 known asteroids and 152 new objects; on April 20 the respective numbers are 49 and 132. We performed this cross matching in October 2014, and hence a `known' objects in this context means one for which, on that date, the MPC possessed an orbit sufficient to determine April 2013 positions with 20 arcsecond accuracy.  Some of these asteroids had in fact been discovered since our observations, but as our analysis software was still under development, we were not able to report them to the Minor Planet Center in time to obtain discovery credit.

\subsection{Night-to-Night Linkages} \label{sec:link}
How many of the previously unknown asteroids we detected on April 19 were also detected on April 20?  To answer this question, we must match the positions and motion rates of asteroids between the two nights.  This is an important problem for all asteroid surveys, but the highly accurate sky motions that come `for free' along with digital tracking detections make it easier. Simply extrapolating forward (or backward) for 24 hours assuming linear motion at constant velocity is often sufficient for matching, but such positions are systematically incorrect by about 30 arcseconds to the west (east) if extrapolating forward (backward) by one night. The source of this offset is the Earth's rotation, which causes the eastward space velocity of an observer on Earth's surface at night to be faster than the orbital velocity that would apply to idealized observations made from the geocenter. 

Much more accurate extrapolations can be obtained using the angular velocity the asteroid would have had if viewed from the geocenter. To obtain an approximate value for this geocentric angular velocity, we simply estimate the component of the angular velocity that is due to the Earth's rotation and subtract it from the measured angular velocity. The angular velocity contributed by Earth's rotation is given by the physical velocity of the observer relative to the geocenter, projected onto the plane perpendicular to the line-of-sight to the asteroid and divided by the asteroid's estimated distance (see Heinze \& Metchev 2015 for a more detailed discussion). The fact that geocentric distance is inversely correlated with total angular velocity allows us to obtain a crude approximation for the distance that is sufficient for our current purpose. We perform a linear fit to distance as a function of inverse angular velocity for known asteroids in our field. The RMS error of these distances is 0.3 AU. This crude distance estimation should not be confused with the far more accurate method of \citet{curves}, but the latter cannot be used at this stage because it requires that measurements of the asteroid have already been linked across two subsequent nights.

As an example of a night-to-night linkage, analyzing our April 19 data reveals a previously unknown $R$ = 21.7 mag asteroid\footnote{One of the 144 two-night discoveries of our survey, it has been designated 2013 HY$_{153}$ by the MPC.} with a motion of -37.36 arcsec/hr east and 10.32 arcsec/hr north, for a total angular velocity of 38.76 arcsec/hr.  Our linear fit based on known asteroids maps this angular velocity to an approximate geocentric distance of 1.55 AU.  At 1.55 AU, the projected rotational velocity of Kitt Peak during our observations (1340.0 km/hr eastward and -4.8 km/hr northward) produces a reflected angular velocity of -1.19 arcsec/hr eastward and 0.00 arcsec/hr northward.  Thus, if measured from the geocenter, the angular velocity of this object would have been $-37.36 + 1.19 = -36.17$ arcsec/hr eastward and 10.32 arcsec/hr northward.  To predict the asteroid's location in our April 20 data, we simply use these velocities to extrapolate linearly forward by the elapsed time between the reference time for the April 19 digital tracking integration and that for the April 20 integration. There is an $R$ = 21.8 magnitude April 20 detection only 2.4 arcseconds away from the resulting position, and its motion vector matches that of the April 19 asteroid to within 0.09 arcsec/hr. 

Using the linkage method illustrated above, we find that 165 of the asteroids detected in our April 19 observations were independently recovered by our detection software in the April 20 data.  Thus 34 asteroids (including 1 known object) are unique to the April 19 data and 16 (including 3 known objects) are unique to the April 20 data. The full count of confirmed asteroids is therefore 215 objects, of which 50 had accurate orbits as of October 2014, and the remaining 165 appeared to be new discoveries of our survey. Of these 165 new asteroids, 33 were automatically detected only in the April 19 data, 13 only in the April 20 data, and 119 were automatically detected in the data sets from both nights. The reality of the 46 single-night objects is not in doubt, since they were confirmed on multiple independent subsets of the data within the night of their discovery. Nevertheless, we attempted to recover them manually by creating check images and pyramid images centered on their extrapolated locations for the nights on which they were not automatically detected. This was highly successful: all but 12 of the single-night asteroids were recovered on a second night. Of the 12 objects not recovered on a second night, in many cases the cause was obvious: the object had moved out of the field or become superimposed on a masked star. In the remaining cases the non-detection may be due to rotational variability or increasing moonlight. 

\subsection{Verification of a Negligible False Positive Rate} \label{sec:falsepos}

We have performed a test to probe whether we have achieved our goal of zero false positives among confirmed objects, and also whether a different set of criteria for confirmation would result in improved sensitivity while maintaining a very low false positive rate. This test consists of a full re-analysis of the April 19 data with the temporal order of the images scrambled by randomly reassigning the image acquisition times. Real asteroids cannot be registered with the timestamps scrambled, so all detections in this analysis must be false positives. At the same time, the images used are identical to those in the real analysis, so the rate and statistics of false positives in the `scrambled' data set should match those in the real data.

Similar to the real data, the scrambled data set exhibits about 40 false positives that correspond to locally noisy regions near cleaner data that caused the automated algorithm to use an inappropriately low sky noise value in estimating the significance of the detection. The bottom left panel in Figure \ref{fig:pyr} shows an example of this. Other false positives appear against a smooth background and accurately mimic a real but low-significance point source. The two strongest examples are shown in the bottom center and bottom right panels of Figure \ref{fig:pyr}. Both have automatically logged significance values of 7.2$\sigma$. We have re-calculated these significance values by creating 301$\times$301 pixel images with our visualization software, allowing us to use a wider background annulus to more accurately measure the noise. The new, more accurate values are 6.9$\sigma$ and 6.5$\sigma$, respectively. One 6.9$\sigma$ false positive in a single night's data is not surprising, and suggests that once obvious false positives have been discounted, the remaining noise in our images is almost perfectly Gaussian.

For comparison to the 6.9$\sigma$ significance of the strongest false positive in our scrambled data set, the identically calculated value for the least-significant one-night asteroid confirmed in our real data (pt2124) is 7.2$\sigma$ --- a level at which a false positive is nine times less likely under Gaussian statistics, and only 0.18 false positives would be expected in our entire survey even if the trial stacks were strictly independent. The pyramid image for pt2124 is much more visually convincing than those for any of the detections in the scrambled data set (see Figure \ref{fig:pyr}). Additionally, the sky motions of pt2124 (-33.80 arcsec/hr east and 14.56 arcsec/hr north) put it near the densest concentration of real asteroids in Figure \ref{fig:drift}, rather than off in one of the empty corners where a truly spurious detection would have an equal probability of falling --- a consideration which reduces its false positive probability by a further factor of at least 2. This asteroid therefore appears to have no more that about a 5\% chance of being a false positive. All of our other confirmed asteroids are detected with far smaller false positive probability than pt2124, due to greater significance and/or consistent detection on both nights.

The scrambled-data analysis supports the reality of all our confirmed asteroids.  However, it also validates the expectation from Section \ref{sec:autothresh} that our analysis probes a sufficiently large number of noise realizations that multiple spurious point sources will appear with significance above 6.5$\sigma$, even with pure Gaussian noise. Indeed, the spurious sources from the scrambled data are indistinguishable from many unconfirmed candidates in the real data set, including those shown in the middle row of Figure \ref{fig:pyr}. While the greater abundance of such detections in the real data relative to the scrambled analysis makes it certain that some of them are genuine asteroids, they cannot be confidently identified. Thus it does not seem that our detection criteria can be significantly relaxed without incurring numerous false positives.

\subsection{Reporting Objects to the MPC} \label{sec:mpc}
We have reported all 215 of our detected asteroids to the MPC: 50 as previously known objects, 153 as two-night discoveries, and 12 as single-night objects. For all but the lowest SNR detections, we reported two positions per night. All of our observations have been accepted and processed. The MPC holds single-night asteroids in a file for later matching, but does not assign discovery designations for them. Thus we still refer to single-night objects by our own temporary designations: e.g., pt2124 in Figures \ref{fig:check} and \ref{fig:pyr}. 

Of our 153 previously unknown two-night objects, the MPC has issued discovery designations for 144. The remaining nine have been matched by the MPC to previously-known objects with orbits too inaccurate for our own matching to have identified them. Our data have resulted in significant improvements to the orbital accuracy of these nine objects. One example is 2013 EW$_{149}$, shown in Figures \ref{fig:check} and \ref{fig:pyr}. It was discovered March 13, 2013 and measured three times over a period of two nights. Apart from these discovery observations, 2013 EW$_{149}$ is known only through our measurements on April 19 and 20. Without our data, no meaningful orbit would be known for this object.

Our 215 detected asteroids can thus be divided into four categories: 50 asteroids with well-known orbits independent of our observations; 9 objects to which our observations contributed significant orbital information; 144 two-night discoveries with designations from the MPC, and 12 single-night objects. Table \ref{tab:asteroids} gives the MPC designation for each of our asteroids, where applicable, and indicates the category to which each object belongs.

\subsection{Measurements of Detected Asteroids} \label{sec:motion}
For each asteroid, we measure the angular velocity, brightness, and celestial coordinates at reference times selected to be near the midpoints of the digital tracking integration on each respective night. For the positions and fluxes, we simply measure the best motion-matched image stacks. Except for the faintest objects, the precision of the positions is better than 0.1 arcsec and is probably limited by astrometric calibration error. We convert our measured asteroid fluxes into magnitudes using a calibration derived from known asteroids in our data, obtaining the magnitudes of these known objects from ephemerides generated by the MPC. The results are likely good to about 0.1 mag except for the faintest objects. Obtaining a more accurate calibration from photometric standard stars would require careful corrections for PSF variations, and the precise magnitudes it would yield are not required at present. The coordinates and magnitudes of all 215 detected asteroids are given in Table \ref{tab:asteroids}.

\begin{figure} 
\includegraphics[scale=0.23]{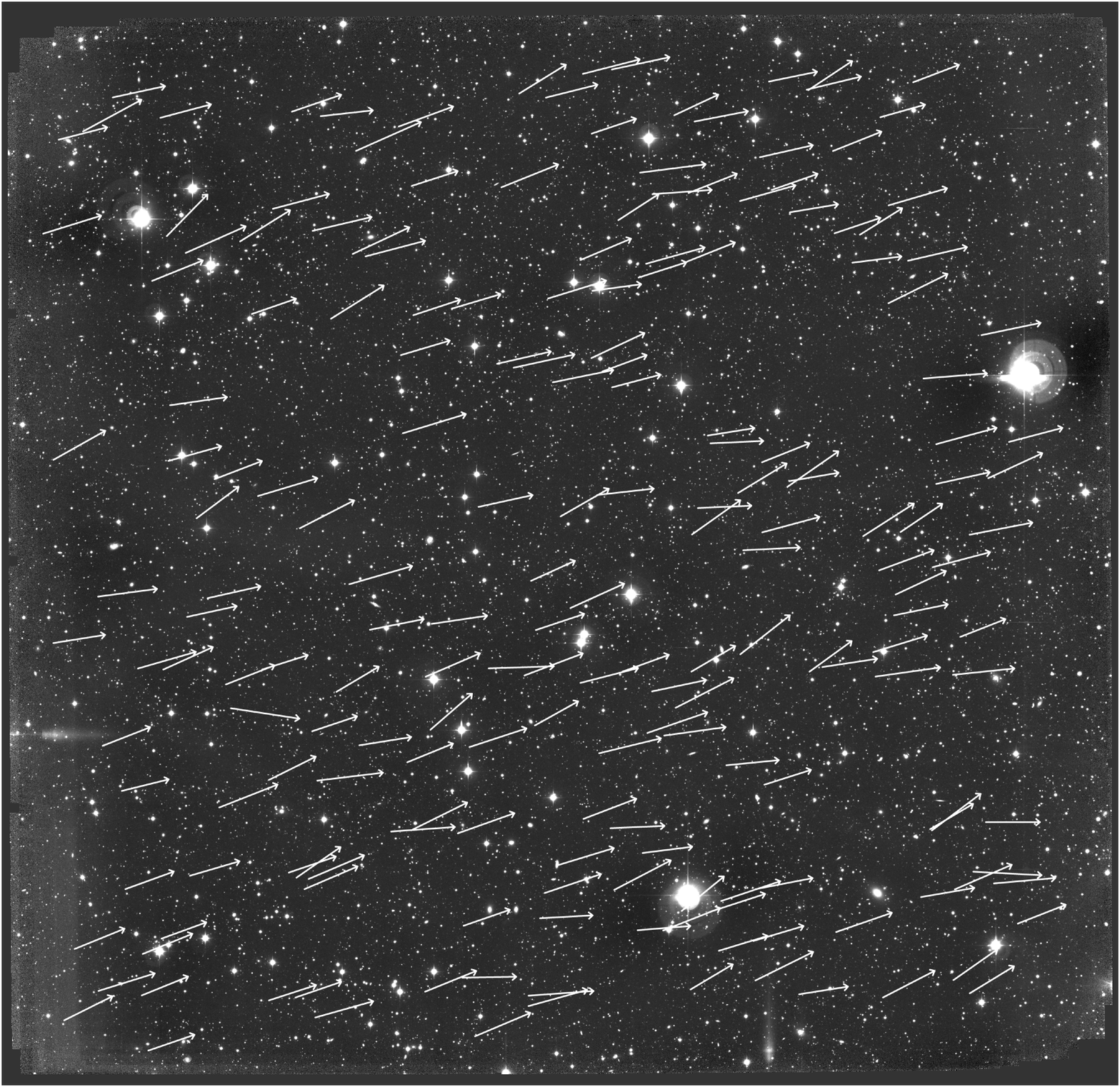}
\caption{The real on-sky tracks of asteroids detected in our April 19 data.  
The background image against which the tracks are plotted is a master star image made by
stacking the same data, and the field of view shown (including the narrow
rim of zero padding) is 1.13$\times$1.10 degrees. Of 199 asteroids, 154 are
new discoveries, and all are verified by consistent detection in multiple
independent subsets of the images.  Their magnitudes extend past $R$ = 23: digital
tracking enables the 0.9-meter telescope to probe a regime previously accessible
only to 4-meter and larger instruments.
\label{fig:tracks}}
\end{figure}

To measure the angular velocities of our asteroids, we use quadratic fits to a grid of digital tracking stacks with motions near the optimal value, as described in Section \ref{sec:checkim}. We calculate the uncertainties based on the residuals from these quadratic fits. The measured tracks of our asteroids relative to the background starfield are shown in Figure \ref{fig:tracks}, and the measured drift rates are plotted and compared to known objects in Figure \ref{fig:drift}.  

\begin{figure} 
\includegraphics[scale=0.8]{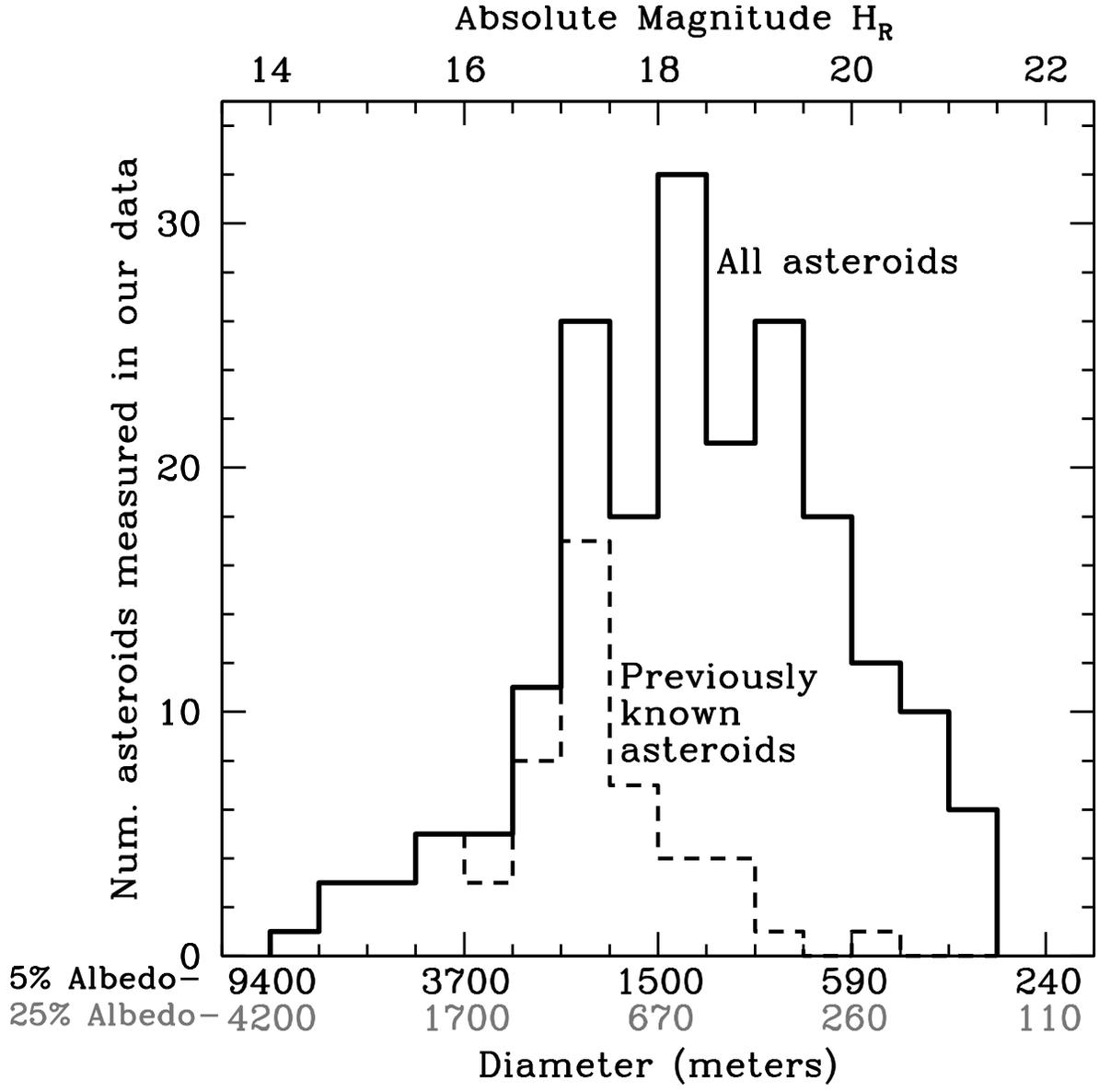}
\caption{Histograms of the absolute magnitudes (and corresponding physical diameters) for all asteroids (solid line) and for previously known objects (dashed line) measured in our data. The current census of the main belt becomes substantially incomplete at a diameter of about 2 km. By contrast, we have detected dozens of new asteroids in the 200-500 meter size range.
\label{fig:distmag}}
\end{figure}

These precisely measured angular velocities allow us to calculate the geocentric distance of each two-night object using Earth rotational velocity reflex as we describe in \citet{curves}. The mean error of our distance determinations is only 1.5\% for known objects. For newly discovered objects, our distances and flux measurements allow us to calculate absolute magnitudes and approximate sizes for the first time. These values are given in Table \ref{tab:asteroids}, and the histogram of absolute magnitudes is shown in Figure \ref{fig:distmag}. The smallest asteroids we have detected have absolute magnitudes $H_R \sim 21.5$, and hence diameters of 130--300 meters depending on their albedo. While asteroids of this size range among the NEOs are routinely detected during close approaches to Earth, our survey is the first to detect them in the main belt with a telescope smaller than 4 meters.

\begin{deluxetable}{lcccccccc}
\tabletypesize{\tiny}
\tablewidth{0pt}
\tablecaption{Asteroids Detected \label{tab:asteroids}}
\tablehead{ & \multicolumn{2}{c}{7.5989 UT Apr 20\tablenotemark{a}} & \multicolumn{2}{c}{7.5669 UT Apr 21\tablenotemark{a}} & & & & \\
\colhead{Object ID} & \colhead{RA} & \colhead {DEC} & \colhead{RA} & \colhead {DEC} & \colhead{R} & \colhead{Dist (AU)} & \colhead{$\mathrm{H}_{\mathrm{R}}$} & \colhead{Diam (km)\tablenotemark{b}}}
\startdata
Hrabe\tablenotemark{c} & 13:57:07.56 & -11:31:06.8 & 13:56:11.76 & -11:27:19.2 & 18.1 & $1.101 \pm 0.010$ & 16.3 & 1.5--3.3 \\
(12231)\tablenotemark{c} & 13:54:38.48 & -11:58:03.5 & 13:53:43.78 & -11:53:50.4 & 18.2 & $2.019 \pm 0.023$ & 14.3 & 3.7--8.2 \\
(123940)\tablenotemark{c} & 13:54:57.46 & -12:19:10.5 & 13:54:08.12 & -12:17:12.3 & 18.3 & $1.901 \pm 0.033$ & 14.6 & 3.2--7.1 \\
(20458)\tablenotemark{c} & 13:55:42.60 & -12:12:08.0 & 13:54:45.43 & -12:04:35.5 & 18.5 & $1.554 \pm 0.013$ & 15.5 & 2.1--4.7 \\
(154215)\tablenotemark{c} & 13:56:30.30 & -12:02:10.2 & 13:55:48.45 & -11:52:57.1 & 18.6 & $1.951 \pm 0.020$ & 14.8 & 2.9--6.5 \\
(82140)\tablenotemark{c} & 13:55:53.89 & -11:55:07.6 & 13:54:59.55 & -11:48:52.6 & 18.7 & $1.771 \pm 0.021$ & 15.2 & 2.4--5.4 \\
(121685)\tablenotemark{c} & 13:56:04.95 & -11:53:36.2 & 13:55:20.69 & -11:48:28.3 & 18.7 & $2.054 \pm 0.022$ & 14.7 & 3.1--6.9 \\
(304006)\tablenotemark{c} & 13:54:17.66 & -11:59:45.4 & 13:53:16.84 & -11:58:08.1 & 18.7 & $0.907 \pm 0.005$ & 17.5 & 0.8--1.9 \\
(93774)\tablenotemark{c} & 13:55:17.52 & -12:16:18.1 & 13:54:28.34 & -12:12:31.1 & 18.9 & $1.869 \pm 0.022$ & 15.2 & 2.4--5.3 \\
(203904)\tablenotemark{c} & 13:56:47.67 & -11:33:29.6 & 13:55:51.46 & -11:26:41.1 & 18.9 & $1.048 \pm 0.008$ & 17.3 & 0.9--2.1 \\
(211984)\tablenotemark{c} & 13:55:50.42 & -11:22:54.4 & 13:54:53.85 & -11:19:12.2 & 18.9 & $1.149 \pm 0.015$ & 17.0 & 1.1--2.4 \\
(162579)\tablenotemark{c} & \nodata & \nodata & 13:56:58.35 & -11:38:29.4 & 19.0 & \nodata & \nodata & \nodata \\
(178125)\tablenotemark{c} & 13:57:45.22 & -11:21:01.7 & 13:56:48.01 & -11:18:10.6 & 19.2 & $1.546 \pm 0.062$ & 16.2 & 1.5--3.4 \\
(251570)\tablenotemark{c} & 13:56:56.95 & -12:20:19.2 & 13:55:58.88 & -12:16:08.6 & 19.3 & $1.205 \pm 0.009$ & 17.2 & 1.0--2.2 \\
(357974)\tablenotemark{c} & 13:55:35.08 & -11:29:06.8 & 13:54:29.56 & -11:26:58.3 & 19.3 & $1.097 \pm 0.009$ & 17.5 & 0.8--1.8 \\
(79454)\tablenotemark{c} & 13:54:51.03 & -11:59:13.6 & 13:53:55.27 & -11:57:06.3 & 19.5 & $2.069 \pm 0.019$ & 15.5 & 2.1--4.7 \\
(187832)\tablenotemark{c} & 13:56:04.07 & -12:02:17.2 & 13:55:20.34 & -11:56:07.0 & 19.5 & $2.231 \pm 0.028$ & 15.2 & 2.5--5.5 \\
2009 FD$_{72}$\tablenotemark{c} & 13:57:16.55 & -12:02:19.2 & 13:56:07.56 & -12:04:58.2 & 19.6 & $1.230 \pm 0.009$ & 17.4 & 0.9--2.0 \\
(257628)\tablenotemark{c} & 13:54:15.51 & -12:18:30.2 & 13:53:30.88 & -12:11:35.8 & 19.6 & $1.978 \pm 0.027$ & 15.8 & 1.9--4.2 \\
(212989)\tablenotemark{c} & 13:54:34.23 & -11:30:49.8 & 13:53:35.48 & -11:26:27.6 & 19.6 & $1.424 \pm 0.012$ & 16.9 & 1.1--2.4 \\
2013 JA$_{53}$\tablenotemark{d} & 13:54:23.25 & -11:52:56.0 & 13:53:26.27 & -11:49:05.7 & 19.6 & $1.039 \pm 0.008$ & 18.0 & 0.7--1.5 \\
pt2132\tablenotemark{f} & 13:54:07.36 & -12:12:29.4 & \nodata & \nodata & 20.0 & \nodata & \nodata & \nodata \\
(361542)\tablenotemark{c} & 13:54:34.12 & -11:36:24.9 & 13:53:34.70 & -11:28:57.1 & 20.2 & $1.398 \pm 0.012$ & 17.6 & 0.8--1.8 \\
(398106)\tablenotemark{c} & 13:54:10.32 & -12:08:55.8 & 13:53:15.19 & -12:08:56.6 & 20.2 & $2.474 \pm 0.044$ & 15.5 & 2.1--4.6 \\
(209656)\tablenotemark{c} & 13:55:58.91 & -11:36:28.9 & 13:55:00.61 & -11:31:55.5 & 20.2 & $1.605 \pm 0.032$ & 17.1 & 1.0--2.2 \\
(152105)\tablenotemark{c} & 13:57:11.68 & -11:59:11.0 & 13:56:25.40 & -11:55:31.3 & 20.3 & $2.417 \pm 0.029$ & 15.7 & 1.9--4.3 \\
(403699)\tablenotemark{c} & 13:55:00.01 & -11:47:54.1 & 13:54:09.99 & -11:45:45.1 & 20.3 & $2.231 \pm 0.024$ & 16.0 & 1.7--3.7 \\
(257171)\tablenotemark{c} & 13:57:09.89 & -12:05:37.9 & 13:56:22.59 & -11:59:41.7 & 20.3 & $1.950 \pm 0.021$ & 16.5 & 1.3--2.9 \\
2005 SC$_{42}$\tablenotemark{c} & 13:55:57.02 & -11:49:26.4 & 13:55:08.49 & -11:42:24.4 & 20.4 & $2.016 \pm 0.024$ & 16.5 & 1.3--3.0 \\
(364373)\tablenotemark{c} & 13:56:43.78 & -11:56:52.8 & 13:55:49.26 & -11:54:10.2 & 20.4 & $1.805 \pm 0.020$ & 16.9 & 1.1--2.5 \\
(172698)\tablenotemark{c} & 13:55:50.71 & -12:08:05.6 & 13:54:57.13 & -12:02:53.9 & 20.4 & $1.946 \pm 0.025$ & 16.6 & 1.3--2.8 \\
(311762)\tablenotemark{c} & 13:54:22.91 & -11:47:58.1 & 13:53:28.80 & -11:44:58.3 & 20.4 & $1.751 \pm 0.017$ & 17.0 & 1.1--2.4 \\
2013 HJ$_{153}$\tablenotemark{e} & 13:55:43.71 & -11:31:27.1 & 13:55:03.36 & -11:24:49.2 & 20.4 & $2.197 \pm 0.022$ & 16.2 & 1.5--3.4 \\
2013 HQ$_{152}$\tablenotemark{e} & 13:55:22.20 & -11:49:43.6 & 13:54:28.26 & -11:49:07.7 & 20.4 & $1.604 \pm 0.014$ & 17.2 & 0.9--2.1 \\
2013 HC$_{155}$\tablenotemark{e} & 13:57:01.16 & -11:50:09.4 & 13:56:07.83 & -11:43:03.5 & 20.4 & $0.965 \pm 0.008$ & 19.0 & 0.4--1.0 \\
2013 HC$_{154}$\tablenotemark{e} & 13:56:12.78 & -11:59:30.0 & 13:55:06.71 & -11:58:57.5 & 20.5 & $1.524 \pm 0.011$ & 17.5 & 0.8--1.9 \\
2013 HK$_{153}$\tablenotemark{e} & 13:55:43.10 & -11:22:42.3 & 13:54:45.15 & -11:19:34.3 & 20.5 & $1.441 \pm 0.019$ & 17.8 & 0.7--1.6 \\
(412517)\tablenotemark{c} & 13:54:36.93 & -11:52:58.2 & 13:53:41.57 & -11:47:50.6 & 20.6 & $1.855 \pm 0.023$ & 17.0 & 1.1--2.3 \\
2013 HL$_{153}$\tablenotemark{e} & 13:55:43.86 & -12:09:13.6 & 13:54:48.74 & -12:08:39.2 & 20.7 & $2.248 \pm 0.031$ & 16.4 & 1.4--3.1 \\
2013 HW$_{154}$\tablenotemark{e} & 13:56:53.62 & -12:00:12.3 & 13:56:10.75 & -11:53:46.7 & 20.7 & $2.161 \pm 0.024$ & 16.5 & 1.3--3.0 \\
2014 RB$_{7}$\tablenotemark{c} & 13:58:04.43 & -11:32:31.5 & 13:57:05.82 & -11:27:53.0 & 20.7 & $1.612 \pm 0.034$ & 17.6 & 0.8--1.8 \\
2005 JW$_{17}$\tablenotemark{c} & 13:56:17.44 & -12:21:14.4 & 13:55:21.70 & -12:15:14.8 & 20.8 & $1.421 \pm 0.014$ & 18.1 & 0.6--1.4 \\
(363018)\tablenotemark{c} & 13:56:07.10 & -11:40:50.3 & 13:55:05.22 & -11:37:26.0 & 20.8 & $1.815 \pm 0.029$ & 17.2 & 0.9--2.1 \\
2013 HD$_{152}$\tablenotemark{e} & 13:55:05.71 & -11:50:47.7 & 13:54:08.88 & -11:46:49.8 & 20.8 & $1.041 \pm 0.010$ & 19.1 & 0.4--0.9 \\
(252335)\tablenotemark{c} & 13:57:41.02 & -12:16:19.7 & 13:56:50.53 & -12:11:10.8 & 20.9 & $2.039 \pm 0.025$ & 17.0 & 1.1--2.4 \\
2013 EJ$_{149}$\tablenotemark{d} & 13:55:15.19 & -12:05:18.5 & 13:54:21.95 & -12:03:58.5 & 20.9 & $1.994 \pm 0.023$ & 17.0 & 1.1--2.4 \\
2013 HT$_{152}$\tablenotemark{e} & 13:55:25.39 & -12:18:20.0 & 13:54:42.73 & -12:11:54.8 & 20.9 & $2.091 \pm 0.024$ & 16.9 & 1.1--2.5 \\
(290287)\tablenotemark{c} & 13:54:03.09 & -12:14:31.2 & 13:53:14.64 & -12:09:45.1 & 20.9 & $1.866 \pm 0.047$ & 17.3 & 0.9--2.1 \\
(266193)\tablenotemark{c} & 13:55:51.80 & -11:34:01.8 & 13:55:00.94 & -11:28:34.3 & 20.9 & $1.825 \pm 0.034$ & 17.3 & 0.9--2.0 \\
2007 TO$_{45}$\tablenotemark{d} & 13:57:28.75 & -11:33:26.0 & 13:56:28.68 & -11:26:38.1 & 20.9 & $1.176 \pm 0.009$ & 18.8 & 0.5--1.0 \\
2013 HS$_{155}$\tablenotemark{e} & 13:57:29.86 & -11:56:05.6 & 13:56:39.91 & -11:53:21.0 & 20.9 & $1.782 \pm 0.020$ & 17.4 & 0.9--1.9 \\
2013 HG$_{154}$\tablenotemark{e} & 13:56:20.80 & -12:18:22.2 & 13:55:25.95 & -12:18:09.3 & 20.9 & $2.091 \pm 0.046$ & 16.8 & 1.2--2.6 \\
2013 HQ$_{156}$\tablenotemark{e} & 13:58:08.94 & -12:05:07.7 & 13:57:12.76 & -11:59:24.1 & 20.9 & \nodata & \nodata & \nodata \\
2013 HX$_{150}$\tablenotemark{e} & 13:54:03.63 & -11:36:45.1 & 13:53:21.99 & -11:27:33.2 & 21.0 & $1.881 \pm 0.045$ & 17.4 & 0.9--2.0 \\
2006 QG$_{123}$\tablenotemark{c} & 13:56:07.54 & -11:23:39.2 & 13:55:20.15 & -11:16:16.6 & 21.1 & $2.384 \pm 0.117$ & 16.5 & 1.3--3.0 \\
pt2166\tablenotemark{f} & \nodata & \nodata & 13:56:29.89 & -11:17:54.3 & 21.1 & \nodata & \nodata & \nodata \\
2013 HO$_{151}$\tablenotemark{e} & 13:54:33.31 & -11:55:59.7 & 13:53:38.87 & -11:53:11.9 & 21.2 & $1.838 \pm 0.025$ & 17.6 & 0.8--1.8 \\
2013 EL$_{135}$\tablenotemark{d} & 13:56:45.77 & -11:27:08.8 & 13:55:40.59 & -11:19:46.5 & 21.2 & $1.184 \pm 0.013$ & 19.1 & 0.4--0.9 \\
2013 HO$_{152}$\tablenotemark{e} & 13:55:18.94 & -12:13:50.1 & 13:54:23.16 & -12:09:13.6 & 21.2 & $1.466 \pm 0.017$ & 18.4 & 0.5--1.2 \\
2013 HH$_{156}$\tablenotemark{e} & 13:58:00.78 & -12:20:14.8 & 13:57:11.96 & -12:14:06.5 & 21.3 & $1.987 \pm 0.048$ & 17.4 & 0.9--1.9 \\
2013 HE$_{154}$\tablenotemark{e} & 13:56:16.70 & -11:49:21.6 & 13:55:21.92 & -11:46:20.5 & 21.3 & $2.090 \pm 0.039$ & 17.3 & 0.9--2.1 \\
2014 QX$_{347}$\tablenotemark{c} & 13:57:41.05 & -11:59:03.3 & 13:56:42.29 & -11:54:47.9 & 21.3 & $1.791 \pm 0.024$ & 17.8 & 0.7--1.6 \\
2005 SC$_{212}$\tablenotemark{c} & 13:57:57.96 & -12:16:01.2 & 13:57:07.44 & -12:10:49.7 & 21.3 & $1.897 \pm 0.028$ & 17.6 & 0.8--1.8 \\
2013 HZ$_{153}$\tablenotemark{e} & 13:56:04.72 & -11:59:18.2 & 13:55:06.00 & -11:53:40.2 & 21.3 & $1.375 \pm 0.016$ & 18.7 & 0.5--1.1 \\
2013 HN$_{155}$\tablenotemark{e} & 13:57:20.33 & -11:59:56.9 & 13:56:31.24 & -11:55:04.5 & 21.3 & $1.929 \pm 0.033$ & 17.5 & 0.8--1.9 \\
2013 HX$_{155}$\tablenotemark{e} & 13:57:36.20 & -11:25:36.7 & 13:56:45.26 & -11:22:11.2 & 21.3 & $1.882 \pm 0.027$ & 17.6 & 0.8--1.8 \\
2013 HB$_{154}$\tablenotemark{e}  & 13:56:12.13 & -11:40:37.8 & 13:55:18.18 & -11:37:30.7 & 21.3 & $1.472 \pm 0.016$ & 18.4 & 0.5--1.2 \\
2013 HY$_{152}$\tablenotemark{e} & 13:55:29.52 & -12:03:23.6 & 13:54:28.42 & -12:01:12.8 & 21.3 & $0.987 \pm 0.009$ & 19.8 & 0.3--0.7 \\
2013 HO$_{153}$\tablenotemark{e} & 13:55:46.85 & -11:48:48.2 & 13:54:50.63 & -11:47:12.7 & 21.3 & $1.167 \pm 0.012$ & 19.2 & 0.4--0.8 \\
2013 HR$_{153}$\tablenotemark{e} & 13:55:50.88 & -11:59:58.9 & 13:54:53.75 & -11:56:10.9 & 21.3 & $1.050 \pm 0.011$ & 19.6 & 0.3--0.7 \\
2013 HN$_{151}$\tablenotemark{e} & 13:54:33.39 & -11:54:20.1 & 13:53:42.55 & -11:48:20.8 & 21.4 & $2.141 \pm 0.039$ & 17.2 & 1.0--2.1 \\
(313002)\tablenotemark{c} & 13:56:59.35 & -12:02:55.6 & 13:56:13.74 & -11:59:05.5 & 21.4 & $2.548 \pm 0.047$ & 16.6 & 1.3--2.8 \\
2005 NG$_{1}$\tablenotemark{c} & 13:55:12.84 & -11:47:57.8 & 13:54:23.57 & -11:40:26.0 & 21.4 & $2.182 \pm 0.050$ & 17.2 & 1.0--2.1 \\
2013 HA$_{155}$\tablenotemark{e} & 13:56:57.33 & -11:32:29.3 & 13:55:57.97 & -11:28:57.0 & 21.5 & $1.474 \pm 0.022$ & 18.7 & 0.5--1.1 \\
2013 HW$_{151}$\tablenotemark{e} & 13:54:49.35 & -11:27:27.0 & 13:54:00.57 & -11:22:30.6 & 21.5 & $1.798 \pm 0.028$ & 18.0 & 0.7--1.5 \\
2013 HH$_{152}$\tablenotemark{e} & 13:55:07.25 & -12:17:40.8 & 13:54:09.87 & -12:10:23.4 & 21.5 & $1.005 \pm 0.010$ & 19.9 & 0.3--0.6 \\
2013 EL$_{150}$\tablenotemark{d} & 13:57:46.50 & -12:06:40.1 & 13:56:58.71 & -12:03:42.5 & 21.6 & $2.491 \pm 0.063$ & 16.9 & 1.1--2.4 \\
2013 HF$_{155}$\tablenotemark{e} & 13:57:03.68 & -11:25:08.1 & 13:56:13.84 & -11:20:48.6 & 21.6 & $2.025 \pm 0.039$ & 17.7 & 0.8--1.7 \\
2013 HG$_{153}$\tablenotemark{e} & 13:55:37.73 & -11:35:17.0 & 13:54:49.50 & -11:31:22.4 & 21.6 & $2.013 \pm 0.035$ & 17.7 & 0.8--1.7 \\
2013 HB$_{156}$\tablenotemark{e} & 13:57:41.68 & -12:18:53.5 & 13:56:54.80 & -12:14:04.2 & 21.6 & $2.010 \pm 0.040$ & 17.7 & 0.8--1.7 \\
2013 HK$_{156}$\tablenotemark{e} & 13:58:02.77 & -11:45:57.0 & 13:57:11.64 & -11:38:29.3 & 21.6 & $1.055 \pm 0.021$ & 20.0 & 0.3--0.6 \\
2013 HH$_{155}$\tablenotemark{e} & 13:57:08.66 & -12:05:25.5 & 13:56:08.04 & -12:06:24.0 & 21.6 & $1.047 \pm 0.013$ & 19.9 & 0.3--0.6 \\
2007 TA$_{195}$\tablenotemark{c} & 13:57:00.64 & -11:49:27.2 & 13:56:05.20 & -11:43:50.6 & 21.6 & $1.737 \pm 0.030$ & 18.2 & 0.6--1.3 \\
pt2159\tablenotemark{f} & \nodata & \nodata & 13:56:07.89 & -11:20:00.7 & 21.6 & \nodata & \nodata & \nodata \\
2013 HU$_{151}$\tablenotemark{e} & 13:54:44.40 & -11:34:40.1 & 13:53:54.44 & -11:33:05.3 & 21.7 & $2.449 \pm 0.050$ & 17.1 & 1.0--2.3 \\
2013 HY$_{153}$\tablenotemark{e} & 13:56:03.69 & -12:19:37.7 & 13:55:04.63 & -12:15:29.8 & 21.7 & $1.479 \pm 0.025$ & 18.9 & 0.4--1.0 \\
2013 HP$_{152}$\tablenotemark{e} & 13:55:19.16 & -11:45:47.3 & 13:54:25.72 & -11:45:08.8 & 21.7 & $1.825 \pm 0.032$ & 18.1 & 0.6--1.4 \\
2013 HH$_{154}$\tablenotemark{e} & 13:56:21.46 & -12:14:44.3 & 13:55:34.85 & -12:10:39.2 & 21.7 & $1.797 \pm 0.030$ & 18.1 & 0.6--1.4 \\
2013 HJ$_{151}$\tablenotemark{e} & 13:54:24.81 & -11:41:45.1 & 13:53:21.65 & -11:40:27.3 & 21.7 & $1.044 \pm 0.017$ & 20.0 & 0.3--0.6 \\
2013 HM$_{155}$\tablenotemark{e} & 13:57:16.50 & -11:32:35.0 & 13:56:26.00 & -11:24:38.1 & 21.8 & $1.916 \pm 0.042$ & 18.1 & 0.6--1.4 \\
2013 EW$_{149}$\tablenotemark{d} & 13:56:01.59 & -12:14:41.7 & 13:55:08.01 & -12:13:56.3 & 21.8 & $2.197 \pm 0.046$ & 17.5 & 0.8--1.8 \\
2013 HB$_{153}$\tablenotemark{e} & 13:55:33.52 & -12:00:36.8 & 13:54:38.64 & -11:57:42.5 & 21.8 & $1.758 \pm 0.031$ & 18.4 & 0.6--1.3 \\
2007 TM$_{173}$\tablenotemark{c} & 13:55:15.62 & -12:13:06.7 & 13:54:15.69 & -12:08:02.3 & 21.8 & $1.638 \pm 0.030$ & 18.7 & 0.5--1.1 \\
2013 HT$_{154}$\tablenotemark{e} & 13:56:44.08 & -11:33:59.5 & 13:55:44.17 & -11:29:59.4 & 21.8 & $1.397 \pm 0.026$ & 19.1 & 0.4--0.9 \\
2013 HC$_{156}$\tablenotemark{e} & 13:57:45.47 & -12:12:27.6 & 13:56:56.62 & -12:07:59.5 & 21.8 & $1.922 \pm 0.039$ & 18.0 & 0.7--1.5 \\
2013 HJ$_{156}$\tablenotemark{e} & 13:58:01.75 & -11:26:58.7 & 13:57:12.88 & -11:23:45.4 & 21.8 & $1.890 \pm 0.071$ & 18.1 & 0.6--1.4 \\
2013 HJ$_{152}$\tablenotemark{e} & 13:55:07.32 & -12:12:40.1 & 13:54:10.93 & -12:09:50.6 & 21.8 & $0.910 \pm 0.009$ & 20.6 & 0.2--0.4 \\
2013 HF$_{151}$\tablenotemark{e} & 13:54:23.57 & -11:48:42.2 & 13:53:25.93 & -11:44:39.3 & 21.8 & $1.856 \pm 0.037$ & 18.2 & 0.6--1.4 \\
(329364)\tablenotemark{c} & 13:55:25.22 & -11:30:00.6 & 13:54:36.11 & -11:25:08.5 & 21.9 & $2.329 \pm 0.060$ & 17.4 & 0.9--2 \\
2011 WH$_{135}$\tablenotemark{c} & 13:56:21.56 & -12:08:58.9 & 13:55:24.31 & -12:03:50.9 & 21.9 & $1.801 \pm 0.038$ & 18.4 & 0.6--1.2 \\
2013 HC$_{153}$\tablenotemark{e} & 13:55:33.91 & -12:02:51.3 & 13:54:33.38 & -11:57:35.3 & 21.9 & $1.371 \pm 0.021$ & 19.3 & 0.4--0.8 \\
2013 HV$_{153}$\tablenotemark{e} & 13:56:00.26 & -11:24:23.9 & 13:55:08.09 & -11:21:06.0 & 21.9 & $1.939 \pm 0.044$ & 18.2 & 0.6--1.4 \\
2013 HP$_{153}$\tablenotemark{e} & 13:55:48.78 & -11:39:53.2 & 13:54:55.22 & -11:33:29.5 & 21.9 & $1.631 \pm 0.042$ & 18.7 & 0.5--1.1 \\
2013 HF$_{153}$\tablenotemark{e} & 13:55:37.25 & -12:15:22.7 & 13:54:43.82 & -12:14:18.2 & 21.9 & $1.774 \pm 0.031$ & 18.5 & 0.5--1.2 \\
2013 HU$_{152}$\tablenotemark{e} & 13:55:25.65 & -11:34:11.5 & 13:54:37.24 & -11:29:19.5 & 21.9 & $2.000 \pm 0.046$ & 18.1 & 0.7--1.5 \\
2013 HZ$_{155}$\tablenotemark{e} & 13:57:38.29 & -11:35:19.6 & 13:56:47.16 & -11:30:05.6 & 21.9 & $1.759 \pm 0.034$ & 18.4 & 0.5--1.2 \\
2013 HE$_{152}$\tablenotemark{e} & 13:55:06.48 & -12:06:11.1 & 13:54:19.45 & -12:02:22.9 & 21.9 & $2.029 \pm 0.047$ & 18.0 & 0.7--1.5 \\
2013 HH$_{153}$\tablenotemark{e} & 13:55:41.87 & -11:59:14.3 & 13:54:54.97 & -11:54:55.4 & 21.9 & $1.881 \pm 0.038$ & 18.2 & 0.6--1.3 \\
2013 HT$_{153}$\tablenotemark{e} & 13:55:57.40 & -11:41:45.6 & 13:54:56.89 & -11:38:36.9 & 21.9 & $1.190 \pm 0.018$ & 19.8 & 0.3--0.7 \\
2013 HA$_{151}$\tablenotemark{e} & 13:54:10.97 & -11:38:53.1 & 13:53:12.43 & -11:35:44.6 & 21.9 & $0.968 \pm 0.015$ & 20.5 & 0.2--0.5 \\
2013 HQ$_{153}$\tablenotemark{e} & 13:55:49.08 & -11:36:22.1 & 13:54:59.41 & -11:34:40.6 & 22.0 & $2.487 \pm 0.086$ & 17.3 & 0.9--2.0 \\
2013 HV$_{155}$\tablenotemark{e} & 13:57:35.91 & -11:31:55.9 & 13:56:55.17 & -11:21:18.7 & 22.0 & $1.955 \pm 0.059$ & 18.2 & 0.6--1.4 \\
2013 HY$_{154}$\tablenotemark{e} & 13:56:56.16 & -11:25:45.1 & 13:56:03.42 & -11:24:15.9 & 22.0 & $1.990 \pm 0.052$ & 18.1 & 0.6--1.4 \\
2013 HM$_{153}$\tablenotemark{e} & 13:55:44.24 & -11:42:03.3 & 13:54:55.46 & -11:38:49.2 & 22.0 & $1.924 \pm 0.040$ & 18.2 & 0.6--1.3 \\
2013 HM$_{154}$\tablenotemark{e} & 13:56:33.84 & -11:37:34.3 & 13:55:46.69 & -11:33:38.7 & 22.0 & $1.732 \pm 0.039$ & 18.6 & 0.5--1.1 \\
2013 HU$_{153}$\tablenotemark{e} & 13:55:59.94 & -12:12:39.1 & 13:55:03.91 & -12:07:47.1 & 22.0 & $1.131 \pm 0.022$ & 20.1 & 0.3--0.6 \\
2013 HY$_{150}$\tablenotemark{e} & 13:54:04.82 & -11:45:19.9 & 13:53:11.24 & -11:41:44.6 & 22.0 & $1.531 \pm 0.077$ & 19.1 & 0.4--0.9 \\
2013 HD$_{154}$\tablenotemark{e} & 13:56:15.68 & -11:49:38.3 & 13:55:29.28 & -11:45:32.4 & 22.0 & $1.830 \pm 0.041$ & 18.4 & 0.5--1.2 \\
2013 HN$_{156}$\tablenotemark{e} & 13:54:03.28 & -12:13:59.3 & 13:53:16.27 & -12:04:08.6 & 22.0 & $1.494 \pm 0.050$ & 19.2 & 0.4--0.9 \\
2013 HG$_{151}$\tablenotemark{e} & 13:54:24.44 & -12:08:17.1 & 13:53:35.37 & -11:58:55.6 & 22.1 & $1.766 \pm 0.042$ & 18.7 & 0.5--1.1 \\
2013 HT$_{151}$\tablenotemark{e} & 13:54:43.42 & -11:32:19.6 & 13:54:00.95 & -11:25:37.7 & 22.1 & $1.990 \pm 0.048$ & 18.2 & 0.6--1.3 \\
2013 HL$_{151}$\tablenotemark{e} & 13:54:29.31 & -11:34:18.6 & 13:53:30.53 & -11:30:50.9 & 22.1 & $1.114 \pm 0.020$ & 20.3 & 0.2--0.5 \\
2013 HZ$_{150}$\tablenotemark{e} & 13:54:09.88 & -11:47:12.0 & 13:53:16.54 & -11:40:51.0 & 22.1 & $1.046 \pm 0.033$ & 20.4 & 0.2--0.5 \\
2013 HO$_{154}$\tablenotemark{e} & 13:56:34.35 & -11:29:43.4 & 13:55:47.68 & -11:25:57.1 & 22.2 & $2.255 \pm 0.072$ & 17.9 & 0.7--1.6 \\
2004 VA$_{38}$\tablenotemark{c} & 13:55:49.82 & -11:26:32.1 & 13:55:05.00 & -11:22:46.5 & 22.2 & $2.567 \pm 0.151$ & 17.4 & 0.9--2.0 \\
2013 HV$_{151}$\tablenotemark{e} & 13:54:49.27 & -11:32:35.0 & 13:54:02.84 & -11:28:41.0 & 22.2 & $2.241 \pm 0.102$ & 17.9 & 0.7--1.6 \\
2013 HM$_{156}$\tablenotemark{e} & 13:58:04.09 & -11:37:09.8 & 13:57:04.94 & -11:32:26.3 & 22.2 & $1.748 \pm 0.101$ & 18.8 & 0.5--1.0 \\
2013 HN$_{154}$\tablenotemark{e} & 13:56:33.27 & -12:08:33.4 & 13:55:36.75 & -12:01:27.7 & 22.3 & $1.582 \pm 0.043$ & 19.3 & 0.4--0.8 \\
2010 WU$_{71}$\tablenotemark{c} & 13:57:14.36 & -11:37:30.5 & 13:56:29.51 & -11:33:37.4 & 22.3 & $2.649 \pm 0.099$ & 17.3 & 0.9--2.0 \\
2013 HM$_{152}$\tablenotemark{e} & 13:55:11.52 & -11:30:36.6 & 13:54:15.56 & -11:26:38.4 & 22.3 & $1.895 \pm 0.068$ & 18.6 & 0.5--1.1 \\
2013 HS$_{152}$\tablenotemark{e} & 13:55:25.03 & -11:58:57.5 & 13:54:40.73 & -11:52:23.2 & 22.3 & $2.027 \pm 0.054$ & 18.3 & 0.6--1.3 \\
2013 HU$_{155}$\tablenotemark{e} & 13:57:34.02 & -11:46:28.5 & 13:56:40.76 & -11:42:46.7 & 22.3 & $1.863 \pm 0.051$ & 18.6 & 0.5--1.1 \\
2013 HP$_{155}$\tablenotemark{e} & 13:57:23.23 & -11:47:28.4 & 13:56:35.55 & -11:42:50.5 & 22.3 & $1.832 \pm 0.058$ & 18.7 & 0.5--1.1 \\
2013 EH$_{150}$\tablenotemark{d} & 13:55:45.70 & -12:04:16.0 & 13:54:43.82 & -12:00:25.0 & 22.3 & $1.205 \pm 0.033$ & 20.2 & 0.2--0.5 \\
2013 HP$_{151}$\tablenotemark{e} & 13:54:36.22 & -12:18:46.5 & 13:53:44.50 & -12:11:59.2 & 22.3 & $1.449 \pm 0.044$ & 19.5 & 0.3--0.7 \\
2013 HT$_{155}$\tablenotemark{e} & 13:57:33.09 & -11:43:17.2 & 13:56:36.24 & -11:41:13.1 & 22.3 & $1.334 \pm 0.031$ & 19.8 & 0.3--0.6 \\
pt2129\tablenotemark{f} & 13:54:11.62 & -12:12:03.4 & \nodata & \nodata & 22.3 & \nodata & \nodata & \nodata \\
2013 HR$_{151}$\tablenotemark{e} & 13:54:38.05 & -11:25:36.4 & 13:53:52.59 & -11:22:25.6 & 22.4 & $2.555 \pm 0.097$ & 17.6 & 0.8--1.8 \\
2013 HL$_{156}$\tablenotemark{e} & 13:58:02.87 & -11:57:45.9 & 13:57:10.24 & -11:55:42.9 & 22.4 & $2.128 \pm 0.383$ & 18.3 & 0.6--1.3 \\
2013 HO$_{155}$\tablenotemark{e} & 13:57:20.78 & -12:07:19.2 & 13:56:21.71 & -12:01:27.1 & 22.4 & $1.675 \pm 0.069$ & 19.1 & 0.4--0.9 \\
2013 EG$_{150}$\tablenotemark{d} & 13:56:36.71 & -11:40:02.0 & 13:55:47.16 & -11:36:19.5 & 22.4 & $2.026 \pm 0.068$ & 18.4 & 0.6--1.2 \\
2013 HE$_{155}$\tablenotemark{e} & 13:57:03.44 & -12:19:07.1 & 13:56:17.10 & -12:15:06.6 & 22.4 & $1.990 \pm 0.069$ & 18.5 & 0.5--1.2 \\
2013 HA$_{156}$\tablenotemark{e} & 13:57:40.07 & -12:22:20.9 & 13:56:52.98 & -12:18:19.3 & 22.4 & $2.066 \pm 0.200$ & 18.3 & 0.6--1.3 \\
2013 HK$_{155}$\tablenotemark{e} & 13:57:10.91 & -12:05:06.9 & 13:56:27.89 & -11:59:07.3 & 22.4 & $2.317 \pm 0.101$ & 18.0 & 0.7--1.5 \\
2014 QA$_{423}$\tablenotemark{c} & 13:54:47.40 & -12:15:01.2 & 13:53:50.85 & -12:10:06.5 & 22.5 & $1.860 \pm 0.054$ & 18.9 & 0.5--1.0 \\
2013 HW$_{153}$\tablenotemark{e} & 13:56:03.16 & -11:56:43.4 & 13:55:14.57 & -11:52:36.5 & 22.5 & $1.756 \pm 0.065$ & 19.0 & 0.4--0.9 \\
pt2122\tablenotemark{f} & 13:54:56.09 & -11:23:26.6 & \nodata & \nodata & 22.5 & \nodata & \nodata & \nodata \\
2013 HE$_{153}$\tablenotemark{e} & 13:55:36.40 & -12:10:36.9 & 13:54:46.17 & -12:09:00.4 & 22.6 & $2.170 \pm 0.062$ & 18.4 & 0.6--1.3 \\
2013 HP$_{154}$\tablenotemark{e} & 13:56:34.39 & -11:44:38.9 & 13:55:31.11 & -11:39:44.8 & 22.6 & $1.274 \pm 0.047$ & 20.3 & 0.2--0.5 \\
2013 HL$_{155}$\tablenotemark{e} & 13:57:10.94 & -11:48:30.6 & 13:56:11.14 & -11:43:57.4 & 22.6 & $1.476 \pm 0.041$ & 19.8 & 0.3--0.7 \\
2013 HL$_{154}$\tablenotemark{e} & 13:56:30.44 & -12:18:36.2 & 13:55:38.82 & -12:13:14.6 & 22.6 & $1.623 \pm 0.053$ & 19.4 & 0.4--0.8 \\
2013 HG$_{155}$\tablenotemark{e} & 13:57:04.92 & -12:11:30.0 & 13:56:17.61 & -12:07:24.6 & 22.6 & $1.823 \pm 0.059$ & 19.0 & 0.4--0.9 \\
2013 HK$_{151}$\tablenotemark{e} & 13:54:29.78 & -11:23:19.6 & 13:53:43.85 & -11:18:55.2 & 22.6 & $1.756 \pm 0.088$ & 19.1 & 0.4--0.9 \\
2013 HC$_{151}$\tablenotemark{e} & 13:54:17.68 & -12:12:15.3 & 13:53:21.01 & -12:05:56.8 & 22.6 & $1.132 \pm 0.051$ & 20.7 & 0.2--0.4 \\
2013 HN$_{152}$\tablenotemark{e} & 13:55:12.12 & -11:57:29.1 & 13:54:23.94 & -11:47:35.9 & 22.6 & $0.898 \pm 0.020$ & 21.4 & 0.1--0.3 \\
2013 HF$_{152}$\tablenotemark{e} & 13:55:07.05 & -11:46:25.3 & 13:54:19.07 & -11:41:35.3 & 22.6 & \nodata & \nodata & \nodata \\
2013 HW$_{152}$\tablenotemark{e} & 13:55:27.14 & -12:01:02.5 & 13:54:29.10 & -11:53:31.2 & 22.7 & $1.317 \pm 0.070$ & 20.3 & 0.2--0.5 \\
2013 HD$_{155}$\tablenotemark{e} & 13:57:03.40 & -12:11:21.4 & 13:56:18.82 & -12:03:36.3 & 22.7 & $2.129 \pm 0.174$ & 18.6 & 0.5--1.1 \\
2013 HD$_{156}$\tablenotemark{e} & 13:57:47.80 & -11:24:26.4 & 13:56:54.87 & -11:21:32.0 & 22.7 & $2.157 \pm 0.135$ & 18.6 & 0.5--1.2 \\
2010 RM$_{180}$\tablenotemark{c} & 13:55:20.65 & -11:45:09.1 & 13:54:32.96 & -11:43:18.2 & 22.7 & $3.096 \pm 0.178$ & 17.1 & 1.0--2.2 \\
2013 HW$_{155}$\tablenotemark{e} & 13:57:35.76 & -11:58:56.2 & 13:56:45.70 & -11:53:13.0 & 22.7 & $1.821 \pm 0.118$ & 19.2 & 0.4--0.9 \\
2013 HX$_{154}$\tablenotemark{e} & 13:56:53.55 & -11:37:13.9 & 13:56:01.34 & -11:28:46.9 & 22.7 & $1.061 \pm 0.032$ & 21.0 & 0.2--0.4 \\
2013 HS$_{153}$\tablenotemark{e} & 13:55:56.67 & -12:11:00.5 & 13:55:01.15 & -12:06:43.1 & 22.7 & $1.119 \pm 0.038$ & 20.8 & 0.2--0.4 \\
pt2156\tablenotemark{f} & \nodata & \nodata & 13:56:06.69 & -12:04:07.8 & 22.7 & \nodata & \nodata & \nodata \\
2013 HL$_{152}$\tablenotemark{e} & 13:55:10.49 & -11:52:17.2 & 13:54:13.06 & -11:51:32.5 & 22.8 & $1.920 \pm 0.103$ & 19.0 & 0.4--0.9 \\
2013 HV$_{152}$\tablenotemark{e} & 13:55:27.61 & -12:13:15.7 & 13:54:44.99 & -12:04:23.8 & 22.8 & $1.798 \pm 0.173$ & 19.2 & 0.4--0.8 \\
2013 HJ$_{154}$\tablenotemark{e} & 13:56:24.06 & -11:37:12.6 & 13:55:34.78 & -11:33:29.7 & 22.8 & $1.969 \pm 0.095$ & 19.0 & 0.4--1 \\
2013 HH$_{151}$\tablenotemark{e} & 13:54:25.63 & -12:08:34.0 & 13:53:42.43 & -12:02:02.7 & 22.8 & $1.955 \pm 0.156$ & 19.0 & 0.4--0.9 \\
2013 HE$_{151}$\tablenotemark{e} & 13:54:22.04 & -11:45:24.0 & 13:53:21.47 & -11:41:15.4 & 22.8 & $1.326 \pm 0.040$ & 20.3 & 0.2--0.5 \\
2013 HZ$_{152}$\tablenotemark{e} & 13:55:30.35 & -12:14:49.5 & 13:54:33.42 & -12:09:05.3 & 22.8 & $1.224 \pm 0.061$ & 20.6 & 0.2--0.4 \\
2013 HS$_{151}$\tablenotemark{e} & 13:54:41.40 & -11:50:29.6 & 13:53:51.12 & -11:42:12.8 & 22.8 & $1.237 \pm 0.038$ & 20.6 & 0.2--0.5 \\
2013 HX$_{153}$\tablenotemark{e} & 13:56:02.85 & -12:19:21.2 & 13:54:57.47 & -12:18:12.4 & 22.8 & $1.365 \pm 0.051$ & 20.2 & 0.2--0.5 \\
2013 HS$_{154}$\tablenotemark{e} & 13:56:37.27 & -12:09:21.3 & 13:55:33.88 & -12:08:08.9 & 22.8 & $0.958 \pm 0.024$ & 21.4 & 0.1--0.3 \\
pt2138\tablenotemark{f} & 13:57:54.43 & -11:25:52.5 & \nodata & \nodata & 22.8 & \nodata & \nodata & \nodata \\
pt2147\tablenotemark{f} & 13:54:13.36 & -11:51:02.5 & \nodata & \nodata & 22.8 & \nodata & \nodata & \nodata \\
2013 HE$_{156}$\tablenotemark{e} & 13:57:51.18 & -12:03:42.0 & 13:57:02.97 & -11:58:57.4 & 22.9 & $2.240 \pm 0.112$ & 18.5 & 0.5--1.2 \\
2013 HF$_{156}$\tablenotemark{e} & 13:57:50.80 & -11:54:57.0 & 13:56:50.85 & -11:52:57.8 & 22.9 & $1.817 \pm 0.149$ & 19.4 & 0.4--0.8 \\
2013 HG$_{156}$\tablenotemark{e} & 13:57:51.20 & -12:18:40.0 & 13:56:54.08 & -12:14:09.9 & 22.9 & $1.892 \pm 0.105$ & 19.1 & 0.4--0.9 \\
2013 HD$_{153}$\tablenotemark{e} & 13:55:34.45 & -11:34:24.9 & 13:54:35.30 & -11:30:18.9 & 22.9 & $1.890 \pm 0.075$ & 19.1 & 0.4--0.9 \\
2013 HX$_{151}$\tablenotemark{e} & 13:54:55.85 & -11:58:48.6 & 13:54:13.51 & -11:50:58.4 & 22.9 & $1.765 \pm 0.103$ & 19.5 & 0.3--0.8 \\
2013 HJ$_{155}$\tablenotemark{e} & 13:57:09.87 & -12:19:18.2 & 13:56:22.23 & -12:15:18.8 & 22.9 & \nodata & \nodata & \nodata \\
2013 HX$_{152}$\tablenotemark{e} & 13:55:29.35 & -11:25:13.9 & 13:54:45.05 & -11:20:00.2 & 22.9 & $2.111 \pm 0.110$ & 18.8 & 0.5--1.0 \\
2013 HQ$_{154}$\tablenotemark{e} & 13:56:35.52 & -12:04:42.1 & 13:55:48.92 & -11:59:55.0 & 23.0 & $2.425 \pm 0.230$ & 18.4 & 0.6--1.2 \\
2013 HR$_{152}$\tablenotemark{e} & 13:55:24.32 & -11:50:19.7 & 13:54:35.81 & -11:41:35.5 & 23.0 & $1.659 \pm 0.108$ & 19.8 & 0.3--0.7 \\
2013 HR$_{154}$\tablenotemark{e} & 13:56:36.48 & -11:25:56.3 & 13:55:41.57 & -11:20:48.5 & 23.0 & $1.730 \pm 0.167$ & 19.7 & 0.3--0.7 \\
2013 HM$_{151}$\tablenotemark{e} & 13:54:33.27 & -11:50:22.8 & 13:53:51.21 & -11:43:12.3 & 23.0 & $1.997 \pm 0.433$ & 19.1 & 0.4--0.9 \\
2013 HY$_{151}$\tablenotemark{e} & 13:54:55.15 & -11:23:54.7 & 13:54:00.28 & -11:20:30.6 & 23.0 & $1.770 \pm 0.105$ & 19.5 & 0.3--0.7 \\
2013 HK$_{152}$\tablenotemark{e} & 13:55:08.72 & -12:15:41.9 & 13:54:18.06 & -12:11:53.1 & 23.0 & $1.674 \pm 0.052$ & 19.7 & 0.3--0.7 \\
2013 HR$_{155}$\tablenotemark{e} & 13:57:29.21 & -12:11:44.5 & 13:56:39.31 & -12:08:02.4 & 23.0 & $1.695 \pm 0.264$ & 19.7 & 0.3--0.7 \\
2013 HF$_{154}$\tablenotemark{e} & 13:56:18.62 & -12:03:46.5 & 13:55:21.57 & -11:58:59.1 & 23.0 & $1.254 \pm 0.084$ & 20.7 & 0.2--0.4 \\
2013 HY$_{155}$\tablenotemark{e} & 13:57:36.92 & -12:15:30.7 & 13:56:51.25 & -12:11:21.9 & 23.0 & $3.247 \pm 0.210$ & 17.2 & 1.0--2.1 \\
2013 HZ$_{151}$\tablenotemark{e} & 13:55:00.07 & -11:31:32.3 & 13:54:11.15 & -11:29:29.2 & 23.0 & \nodata & \nodata & \nodata \\
pt2139\tablenotemark{f} & 13:54:26.57 & -12:13:12.5 & \nodata & \nodata & 23.0 & \nodata & \nodata & \nodata \\
pt2140\tablenotemark{f} & 13:57:28.41 & -11:49:26.6 & \nodata & \nodata & 23.0 & \nodata & \nodata & \nodata \\
2013 HC$_{152 }$\tablenotemark{e} & 13:55:05.29 & -11:23:31.1 & 13:54:16.42 & -11:20:59.5 & 23.0 & $2.613 \pm 0.202$ & 18.1 & 0.6--1.4 \\
2013 HB$_{151}$\tablenotemark{e} & 13:54:17.93 & -11:57:05.5 & 13:53:32.50 & -11:52:32.4 & 23.0 & $1.740 \pm 0.365$ & 19.6 & 0.3--0.7 \\
2014 SN$_{155}$\tablenotemark{d} & 13:56:29.85 & -11:59:20.3 & 13:55:34.53 & -11:53:18.4 & 23.0 & $2.153 \pm 0.148$ & 18.7 & 0.5--1.1 \\
2013 HU$_{154}$\tablenotemark{e} & 13:56:46.71 & -12:03:59.6 & 13:55:54.22 & -12:01:53.2 & 23.0 & $1.795 \pm 0.215$ & 19.5 & 0.3--0.8 \\
2013 HN$_{153}$\tablenotemark{e} & 13:55:47.18 & -11:40:58.5 & 13:55:01.88 & -11:36:47.4 & 23.1 & \nodata & \nodata & \nodata \\
2013 HD$_{151}$\tablenotemark{e} & 13:54:19.64 & -12:17:35.3 & 13:53:35.22 & -12:09:42.7 & 23.1 & $2.112 \pm 0.373$ & 18.9 & 0.4--1.0 \\
2013 HB$_{152}$\tablenotemark{e} & 13:55:04.86 & -11:29:55.8 & 13:54:18.64 & -11:25:47.8 & 23.1 & $2.022 \pm 0.111$ & 19.1 & 0.4--0.9 \\
2013 HA$_{152}$\tablenotemark{e} & 13:55:00.25 & -11:47:23.8 & 13:54:09.56 & -11:38:21.2 & 23.1 & $1.258 \pm 0.038$ & 20.8 & 0.2--0.4 \\
2013 HA$_{153}$\tablenotemark{e} & 13:55:32.82 & -11:30:33.2 & 13:54:33.72 & -11:29:28.3 & 23.1 & $1.153 \pm 0.045$ & 21.1 & 0.2--0.4 \\
2013 HQ$_{151}$\tablenotemark{e} & 13:54:36.51 & -11:59:46.8 & 13:53:31.39 & -11:57:26.9 & 23.1 & $1.113 \pm 0.037$ & 21.2 & 0.2--0.3 \\
2013 HA$_{154}$\tablenotemark{e} & 13:56:10.65 & -11:29:33.1 & 13:55:13.52 & -11:23:42.3 & 23.1 & $1.512 \pm 0.236$ & 20.2 & 0.2--0.5 \\
2013 HV$_{154}$\tablenotemark{e} & 13:56:47.77 & -11:53:52.5 & 13:55:44.63 & -11:49:13.1 & 23.2 & $1.343 \pm 0.093$ & 20.7 & 0.2--0.4 \\
2013 HK$_{154}$\tablenotemark{e} & 13:56:28.18 & -11:56:36.0 & 13:55:32.34 & -11:54:28.7 & 23.2 & $1.483 \pm 0.088$ & 20.3 & 0.2--0.5 \\
2013 HB$_{155}$\tablenotemark{e} & 13:56:58.87 & -12:11:39.2 & 13:56:03.76 & -12:05:34.0 & 23.2 & $1.197 \pm 0.062$ & 21.1 & 0.2--0.4 \\
2013 HQ$_{155}$\tablenotemark{e} & 13:57:24.31 & -11:54:55.2 & 13:56:30.76 & -11:51:50.6 & 23.2 & $1.768 \pm 0.136$ & 19.7 & 0.3--0.7 \\
2013 HG$_{152}$\tablenotemark{e} & 13:55:06.92 & -11:28:03.9 & 13:54:13.46 & -11:25:01.8 & 23.2 & $1.690 \pm 0.089$ & 19.9 & 0.3--0.6 \\
pt2124\tablenotemark{f} & 13:57:00.21 & -12:12:17.9 & \nodata & \nodata & 23.3 & \nodata & \nodata & \nodata \\
2013 HZ$_{154}$\tablenotemark{e} & 13:56:55.40 & -12:06:09.2 & 13:55:49.12 & -12:04:04.1 & 23.3 & $1.313 \pm 0.052$ & 20.9 & 0.2--0.4 \\
pt2150\tablenotemark{f} & 13:55:23.10 & -11:25:56.6 & \nodata & \nodata & 23.3 & \nodata & \nodata & \nodata \\
2013 HO$_{156}$\tablenotemark{e} & 13:55:23.01 & -11:39:47.1 & 13:54:26.10 & -11:39:11.5 & 23.4 & $2.073 \pm 0.120$ & 19.2 & 0.4--0.9 \\
2013 HP$_{156}$\tablenotemark{e} & 13:57:20.99 & -11:56:28.1 & 13:56:37.64 & -11:50:10.9 & 23.4 & $2.110 \pm 0.462$ & 19.3 & 0.4--0.8 \\
\enddata
\tablenotetext{a}{UT dates are given here. The Kitt Peak evening dates quoted throughout the
text are one day earlier.}
\tablenotetext{b}{The smaller diameter value corresponds to an albedo of 25\%; the larger diameter is for 5\% albedo.}
\tablenotetext{c}{Object with a well-known orbit independent of our observations.}
\tablenotetext{d}{Object discovered independent of our observations, but for which we have made an important contribution to orbit determination.}
\tablenotetext{e}{Two-night object discovered by our survey and assigned an official designation by the MPC.}
\tablenotetext{f}{One-night object discovered by our survey and known only by our temporary designation.}
\end{deluxetable}

As illustrated by Figure \ref{fig:distmag}, we have detected a large number of asteroids smaller than 500 meters, a size regime where the current census of the main belt has very low completeness. It would be possible to probe the statistics of asteroid sizes and absolute magnitudes in this regime using our data, but this would require an extensive completeness analysis that is beyond the scope of the current work. It would also be of limited scientific value, since the statistics of MBAs in this regime have been probed by \citet{Gladman2009} and \citet{Yoshida2007} using 4--8 meter telescopes. The fact that we have reached a comparable regime using only an 0.9-meter telescope shows the power of digital tracking. A digital tracking survey on a 4--8 meter telescope would easily probe fainter asteroids than have ever previously been detected, and would open up a new regime for statistical studies of the main belt --- especially when combined with the distance determination method we describe in \citet{curves}.

\section{Comparison of Digital Tracking and Conventional Methods} \label{sec:maincomp}

\subsection{A Specific Case: 0.9-meter Digital Tracking and a 4-meter Conventional Survey}

To compare the sensitivity of our digital tracking analysis to that of conventional asteroid surveys, we use only the set of asteroids we have confirmed as genuine. Each was initially detected automatically as a $\ge 7\sigma$ source in one of our digital tracking trial stacks. They have been further subjected to manual evaluations, using conservative criteria as detailed in Section \ref{sec:mancheck}. The final result is that some probably-real objects are rejected, but no (or no more than one) false positive could plausibly exist in the final list. Thanks to this conservatism, our asteroids could be used, e.g., to probe main belt size statistics without followup observations to confirm their reality. This is an important consideration because most of our new asteroids are faint enough that followup from other observatories cannot reasonably be expected, and this will be even more true of asteroids detected in future digital tracking surveys with larger telescopes.

Within the $\sim$1 deg$^2$ field of our observations, a total of 211 asteroids are confirmed in our April 19 data, including twelve that were recovered based on an initial detection in the April 20 images. The faintest of these 211 asteroids have $R$-band magnitudes of 23.4. The SKADS asteroid survey of \citet{Gladman2009} used the 4-meter Mayall Telescope at Kitt Peak without digital tracking to reach a limiting magnitude of $R=23.5$ under much more favorable lunar conditions (lunar ages 2--6 days versus 10--11 days for us), and found an on-sky density of 210 asteroids/deg$^2$ at magnitudes brighter than $R=23.0$. The $R=23.5$ limiting magnitude of \citet{Gladman2009} is not directly comparable to our value of $R=23.4$, since the former corresponds to roughly 50\% completeness while the latter represents the faintest objects detected at any completeness. A robust derivation of our own 50\% completeness limit, while feasible, requires a statistical analysis beyond the scope of the present work.  Nevertheless, the similarity in on-sky number densities between our observations and those of \citet{Gladman2009} suggests our true limiting magnitude at 50\% completeness is close to $R=23.0$. Thus, while our observations are not quite as sensitive as those of the SKADS survey, we have probed a similar regime using a telescope with less than one-sixteenth the collecting area. 

\subsection{More General Advantages and Disadvantages of Digital Tracking} \label{sec:comp}

Some disadvantages of digital tracking should be considered when deciding whether to apply it or conventional methods to a particular science case. Digital tracking requires much longer times spent on each field, so a conventional survey can cover a larger angular area each night. Long digital tracking integrations average the seeing throughout the night; by contrast, conventional surveys sometimes serendipitously obtain very sharp images on which unusually faint objects can be detected. 

Initial detection thresholds for digital tracking stacks can never be much below 7$\sigma$, or the false positive rate based on pure Gaussian statistics will be non-negligible (Section \ref{sec:autothresh}). By contrast, conventional surveys that take at least three images per field can use lower thresholds for detecting objects on individual images because real objects will distinguish themselves from false positives by appearing in subsequent images along a consistent motion vector. In principle, the Gaussian false positive probability for a sequence of three independent 4$\sigma$ detections is lower than that for a single detection at 7$\sigma$. This advantage of the conventional method is partially balanced by the fact that cosmic rays, ghosts, and other artifacts typically enforce a higher detection threshold by making the statistics of individual images much more non-Gaussian than those of our trial stacks, which are made through a clipped median combine of over one hundred individual images. As an example, Pan-STARRS currently uses a 5$\sigma$ threshold \citep{Denneau2013}.

The $\sqrt{t/\tau_M}$ sensitivity advantage for digital tracking versus conventional searches applies to detections at the same significance, and thus the true advantage of digital tracking is less than this factor when compared to conventional searches that use at least three images per field and apply a well-optimized moving object detection methodology like that of \citet{Denneau2013}. However, this does not contradict our claim of a factor of ten sensitivity increase for digital tracking.  For example, under good conditions it is easy to obtain a digital tracking integration consisting of 180 individual exposures of length $\tau_M$ (Section \ref{sec:comp3}), so that $\sqrt{t/\tau_M} = \sqrt{180} = 13.4$. An object detected at 7$\sigma$ in such an integration would therefore be 13.4 times fainter than an object detected at the same significance on a single image, and it would be 9.6 times fainter even than an object detected at 5$\sigma$ on a single frame. Thus, while the true advantage of the digital tracking survey in terms of discovery sensitivity is less than the nominal value of $\sqrt{180}$, it remains approximately a factor of ten.

The biggest advantage of digital tracking is that it enables the detection of asteroids and Kuiper Belt objects that are simply out of reach by any other method: too faint and/or moving too quickly. It thus allows statistical analyses of populations of extremely small objects, and could also extend the reach of surveys aimed at retiring terrestrial impact risk from potentially hazardous NEOs. Another significant advantage is that with digital tracking, every detection comes with a precise measurement of the asteroid's motion, which can aid in determining distances \citep{curves} and orbital information even for extremely faint objects unlikely to be recovered over longer timescales of weeks to years.

\section{Conclusion} \label{sec:conc}
We have described the technique of digital tracking, focusing on its application to searches for faint asteroids and Kuiper Belt objects by stacking tens to hundreds of images from the current generation of large-format CCD mosaic imagers. Digital tracking is suitable for observations of near-Earth objects (NEOs), main belt asteroids (MBAs), and Kuiper Belt objects (KBOs). For all three classes of objects, it typically yields a factor of $\ge 10$ increase in the sensitivity of a telescope to faint moving objects, as compared to conventional techniques.
 
The linear, constant velocity approximation for objects' motions greatly simplifies digital tracking computations, and is sufficient for all-night integrations on MBAs; two-hour integrations targeting NEOs transiting the observer's meridian at 0.1 AU geocentric distance; and 12-night integrations centered on opposition and targeting KBOs. Integrations must be reduced to one hour for NEOs at a distance of 0.1 AU when they are observed far from the meridian (i.e. hour angle far from zero), and to three nights for KBOs one month from opposition.  For MBAs, multi-night integrations are never possible using the linear constant velocity approximation, but all-night (8--9 hour) integrations are possible even when the objects are far from opposition. While previous digital tracking observations of outer Solar System objects have been very valuable scientifically, and \citet{Zhai2014} have made a remarkable detection of a very small asteroid during a close approach to Earth, no previous work has exploited the full potential of digital tracking with large format CCD imagers.

The computational challenge of digital tracking depends on the size of the region of angular motion phase space that must be searched, and the fineness of the required sampling over this region.  We have determined the appropriate parameters for digital tracking surveys targeting various classes of objects, and demonstrated that all of them are computationally tractable. Rapid analysis (hours to days) can be achieved with a supercomputer. In many cases, a high-end desktop workstation is sufficient if a few weeks are available to analyze each night's data, and for a risk-retirement survey requiring faster processing a small cluster may be sufficient.

We have carried out the first digital tracking survey to target asteroids using a large-format CCD imager. Using an 0.9-meter telescope, we have detected MBAs fainter than 23rd magnitude in the $R$ band, thus probing a regime previously explored only with 4-meter and larger telescopes. Within a field of view of approximately 1 square degree, observed with all-night digital tracking integrations on the nights of April 19 and 20, 2013, we have detected a total of 215 asteroids (see Table \ref{tab:asteroids}), of which only 59 were previously known. All 156 of the previously unknown asteroids were manually checked and confirmed to be real based on significant detection in multiple independent subsets of the data.

Among our 215 detected asteroids, 197 were measured on both April 19 and April 20 with sufficient precision that we could derive meaningful distances for them using Earth rotational reflex velocities as described in \citet{curves}. The resulting precise distances allow us to calculate absolute magnitudes and hence approximate diameters for our newly discovered asteroids. Our faintest objects have $H_R \sim 21.5$ mag and hence diameters of 130--300 meters depending on their albedo. While the current census of the main belt becomes substantially incomplete at a diameter of about 2 km, we have detected dozens of new asteroids in the 200--500 meter size range with an 0.9-meter telescope.

To conclude, we have successfully employed digital tracking to detect large numbers of previously unknown asteroids with a very small telescope. We have described the enormous potential of digital tracking, and demonstrated solutions to all significant problems with its large-scale implementation. Our methodology is completely scalable up to the largest telescopes in existence, and will allow them to detect fainter asteroids than have ever previously been imaged. Furthermore, the precise motion measurements that are intrinsically included in digital tracking detections allow accurate geocentric distances to be obtained for any asteroids observed on two nights \citep{curves}. While asteroidal science cases remain for which digital tracking is not the optimal solution, the time is ripe for its widespread deployment in the next generation of asteroid surveys.

\section{Acknowledgments} 
Based on observations at Kitt Peak National Observatory, National Optical Astronomy Observatory (NOAO Prop. ID: 2013A-0501; PI: Aren Heinze), which is operated by the Association of Universities for Research in Astronomy (AURA) under a cooperative agreement with the National Science Foundation.

This publication makes use of the SIMBAD online database,
operated at CDS, Strasbourg, France, and the VizieR online database (see \citet{vizier}).

This publication makes use of data products from the Two Micron All Sky Survey, 
which is a joint project of the University of Massachusetts and the Infrared Processing
and Analysis Center/California Institute of Technology, funded by the National Aeronautics
and Space Administration and the National Science Foundation.

We have also made extensive use of information and code from \citet{nrc}. 
We have used digitized images from the Palomar Sky Survey 
(available from \url{http://stdatu.stsci.edu/cgi-bin/dss\_form}),
 which were produced at the Space 
Telescope Science Institute under U.S. Government grant NAG W-2166. 
The images of these surveys are based on photographic data obtained 
using the Oschin Schmidt Telescope on Palomar Mountain and the UK Schmidt Telescope.

Facilities: \facility{0.9m WIYN}


\begin{thebibliography}{}
\bibitem[Alard \& Lupton(1998)]{Alard1998} Alard, C. \& Lupton, R. H. 1998, \apj, 503, 325
\bibitem[Allen et al.(2001)]{Allen2001} Allen, R. L., Bernstein, G. M., \& Malhotra, R. 2001, \apj, 549, L241
\bibitem[Bernstein et al.(2004)]{Bernstein2004} Bernstein, G. M., Trilling, D. E., Allen, R. L., Brown, M. E., Holman, M., \& Malhotra, R. 2004, \aj, 128, 1364
\bibitem[Chiang \& Brown(1999)]{Chiang1999} Chiang, E. I. \& Brown, M. E. 1999,
\aj, 118, 1411
\bibitem[Cochran et al.(1995)]{Cochran1995} Cochran, A. L., Levison, H. F., Stern, S. A., \& Duncan, M. J. 1995, \apj, 455, 342
\bibitem[Denneau et al.(2013)]{Denneau2013} Denneau, L, Jedicke, R., Grav, T., et al. 2013, \pasp, 125, 357
\bibitem[Flaugher et al.(2012)]{DECam} Flaugher, B. L., Abbot, T. M. C., Angstadt, R. et al. 2012, Proc SPIE, 8446, 844611
\bibitem[Fraser et al.(2008)]{Fraser2008} Fraser, W. C., Kavelaars, J. J., Holman, M. J., Pritchet, C. J., Gladman, B. J., Grav, T., Jones, R. L., MacWilliams, J., \& Petit, J.-M. 2008, Icarus, 195, 827
\bibitem[Fraser \& Kavelaars(2009)]{Fraser2009} Fraser, W. C. \& Kavelaars, J. J. 2009, \aj, 137, 72
%
\bibitem[Fuentes et al.(2009)]{Fuentes2009} Fuentes, C. I., George, M. R., \& Holman, M. J. 2009, \apj, 696, 91
\bibitem[Gehrels \& Jedicke(1996)]{Gehrels1996} Gehrels, T. \& Jedicke, R. 1996, Earth, Moon, and Planets, 72, 233
\bibitem[Gladman et al.(1998)]{Gladman1998} Gladman, B., Kavelaars, J. J., Nicholson, P. D., Loredo, T. J., \& Burns, J. A. 1998, \aj, 116, 2042
\bibitem[Gladman et al.(2001)]{Gladman2001} Gladman, B., Kavelaars, J. J., Petit, J.-M., Morbidelli, A., Holman, M. J., \& Loredo, T. 2001, \aj, 122, 1051
\bibitem[Gladman et al.(2009)]{Gladman2009} Gladman, B. J., Davis, D. R., Neese, C., Jedicke, R., Williams, G., Kavelaars, J. J., Petit, J-M., Scholl, H., Holman, M., Warrington, B., Esquerdo, G., \& Tricarico, P. 2009, Icarus, 202, 104
\bibitem[Gural et al.(2005)]{Gural2005} Gural, P. S., Larsen, J. A., \& Gleason, A. E. 2005, \aj, 130, 1951
\bibitem[Heinze \& Metchev(2015)]{curves} Heinze, A. N. \& Metchev, S. 2015, \aj, submitted
\bibitem[Helin et al.(1997)]{Helin1997} Helin, E., Pravdo, S., Rabinowitz, D. \& Lawrence, K. 1997, Ann. N. Y. Acad. Sci., 822, 6
\bibitem[Holman et al.(2004)]{Holman2004} Holman, M. J., Kavelaars, J. J., Grav, T., Gladman, B. J., Fraser, W. C., Milisavljevic, D., Nicholson, P. D., Burns, J. A., Carruba, V., Petit, J.-M., Rousselot, P., Mousis, O., Marsden, B. G., \& Jacobson, R. A. 2004, Nature, 430, 865
\bibitem[Kavelaars et al.(2004)]{Kavelaars2004} Kavelaars, J. J., Holman, M. J., Grav, T., Milisavljevic, D., Fraser, W., Gladman, B. J., Petit, J.-M., Rousselot, P., Mousis, O., \& Nicholson, P. D. 2004, Icarus, 169, 474
\bibitem[Larson (2007)]{Larson2007} Larson, S. 2007, in IAU Symp. 236, Near Earth Objects, our Celestial Neighbors: Opportunity and Risk, ed. G. B. Valsecchi, \& D.
Vokrouhlick\'{y}, (Cambridge: Cambridge University Press), 323
\bibitem[Luu \& Jewitt(1998)]{Luu1998} Luu, J. X., Jewitt, D. C. 1998, \apj, 502, L91
\bibitem[Ochsenbein et al.(2000)]{vizier} Ochsenbein, F., Bauer, P. \& Marcout, J. 2000, \apjs, 143, 23O
\bibitem[Mainzer et al.(2012)]{mainzer2012} Mainzer, A., Grav, T., Masiero, J., Bauer, J., Cutri, R. M., McMillan, R. S., Nugent, C. R., Tholen, D., Walker, R., \& Wright, E. L. 2012, \apj, 760, L12
\bibitem[Parker \& Kavelaars (2010)]{Parker2010} Parker, A. H. \& Kavelaars, J. J. 2010, \pasp, 122, 549
\bibitem[Pravdo et al.(1999)]{Pravdo1999} Pravdo, S. H., Rabinowitz, D. L., Helin, E. F., Lawrence, K. J., Bambery, R. J., Clark, C. C., Groom, S. L., Levin, S., Lorre, J., Shaklan, S. B., Kervin, P., Africano, J. A., Sydney, P., \& Vicki, S. 1999, \aj, 117, 1616
\bibitem[Press et al.(1992)]{nrc} Press, W. H., Teukolsky, S.A., Vetterling, W. T., \& Flannery, B. P. 1992, Numerical Recipes in C (Second Edition; New York, NY: Cambridge University Press)
\bibitem[Stokes et al.(2000)]{Stokes2000} Stokes, G. H., Evans, J. B., Viggh, H. E. M., Shelly, F. C., \& Pearce, E. C. 2000, Icarus, 148, 21
\bibitem[Shao et al.(2014)]{Shao2014} Shao, M., Nemati, B., Zhai, C., Turyshev, S. G., Sandhu, J., Hallin, G. W., \& Harding, L. K. 2014, \apj, 782, 1
\bibitem[Tyson et al.(1992)]{Tyson1992}  Tyson, J. A., Guhathakurta, P, Bernstein, G., \& Hut, P. 1992, BAAS, 24, 1127
\bibitem[van Dokkum(2001)]{lacos} van Dokkum, P. 2001, \pasp, 113, 1420
\bibitem[Yoshida \& Nakamura (2007)]{Yoshida2007} Yoshida, F. \&  Nakamura, T. 2007, P\&SS, 55, 1113
\bibitem[Zhai et al.(2014)]{Zhai2014} Zhai, C., Shao, M., Nemati, B., Werne, T., Zhou, H., Turyshev, S. G., Sandhu, J., Hallinan, G., \& Harding, L. K. 2014, \apj, 792, 60

\end{thebibliography}
\end{document}